\documentclass[10pt]{article} 
\usepackage[preprint]{tmlr}

\makeatletter
\if@preprint
  
\fi
\makeatother

\usepackage[T1]{fontenc}

\usepackage{hyperref}
\usepackage{url}

\makeatletter
\if@accepted
  \newcommand{\coderepo}{\url{https://github.com/ymoslem/CRE-Router}}
\else
  \newcommand{\coderepo}{\url{https://anonymous.4open.science/r/CRE-Router}}
\fi
\makeatother

\usepackage{graphicx}
\usepackage{siunitx}
\usepackage{subcaption}

\usepackage{tikz}
\usetikzlibrary{fit,positioning,shapes.geometric,arrows.meta}

\usepackage{amsmath,amssymb}

\DeclareMathOperator*{\argmax}{arg\,max}

\usepackage{booktabs}
\usepackage{adjustbox}
\usepackage{multirow}
\usepackage[normalem]{ulem}

\usepackage{xcolor}
\usepackage{colortbl}
\usepackage{array}
\usepackage{ragged2e}

\usepackage[all]{nowidow}
\usepackage{float}
\usepackage{needspace}

\usepackage{enumitem}

\title{Cluster, Route, Escalate:\\Cascaded Framework for Cost-Aware LLM Serving}

\author{%
  \textbf{\normalsize Yasmin Moslem}{\small $^{1}$\normalsize$^{*}$} \hspace{4pt}
  \textbf{\normalsize Magdalena Kacmajor}{\small $^{1}$} \hspace{4pt}
  \textbf{\normalsize David Lynch}{\small $^{2}$} \\
  \textbf{\normalsize Zhuangzhuang Nie}{\small $^{2}$} \hspace{4pt}
  \textbf{\normalsize Patricia Buffini}{\small $^{1}$} \hspace{4pt}
  \textbf{\normalsize John D. Kelleher}{\small $^{1}$\normalsize$^{*}$} \\[10pt]
  {\normalsize \textit{Extended Team:}} \hspace{2pt}
  \textbf{\normalsize Vasudevan Nedumpozhimana}{\small $^{1}$} \hspace{2.5pt}
  \textbf{\normalsize Ammar Abbas}{\small $^{1}$} \hspace{2.5pt}
  \textbf{\normalsize Solmaz Panahi}{\small $^{1}$} \\
  \textbf{\normalsize Alexandros Agapitos}{\small $^{2}$} \hspace{2.5pt}
  \textbf{\normalsize Aleksandar Milenovic}{\small $^{2}$} \hspace{2.5pt}
  \textbf{\normalsize Hongmeng Song}{\small $^{2}$} \hspace{2.5pt}
  \textbf{\normalsize Yucheng Shi}{\small $^{2}$} \hspace{2.5pt}
  \textbf{\normalsize Yue Pan}{\small $^{2}$} \\[10pt]
  {\normalsize\mdseries\itshape $^{1}$ADAPT Centre, Trinity College Dublin \hspace{6pt}
   $^{2}$Huawei Research} \\
  {\mdseries\textit{\scriptsize $^{*}$Corresponding Authors:}
  \texttt{\scriptsize yasmin.moslem, john.kelleher~\{at\}~adaptcentre.ie}}
}

\begin{document}

\maketitle

\begin{abstract}
Serving open-weight language models on premises forces a trade-off between
accuracy and computational cost.
Hosting models of different sizes lets easy queries go to a small,
efficient model and hard ones to a large model.
We present CRE-Router (Cluster, Route, Escalate), a two-stage method that
decides which of these models answers each query.
It clusters queries and routes each cluster to the model that best trades error
against cost, under a budget on the time per output token (TPOT) or the
end-to-end latency per request (E2EL).
It then either accepts the answer or, when a quality estimator rejects it,
escalates the query to the strong model.
Routing decides before generation, so a hard query goes straight to the strong
model; escalation decides after, so it can catch the errors routing cannot
predict.
It needs no annotation beyond task-correctness labels.
We evaluate CRE-Router on TeleMath, TeleQnA and AIME, with pools of two to nine models
served on A100 GPUs.
CRE-Router is more accurate than the efficient models in its pool and comparable
to the strong model it escalates to, at a lower cost per token.
We also find that the choice of cost term decides which models serve the
workload, that pricing each model per cluster shortens the wait per request, and
that a change of hardware changes the best routing.
\end{abstract}

\section{Introduction}
\label{sec:intro}

Open-weight language models now span a wide range of sizes and capabilities,
which encourages operators to deploy several models on their own hardware to
meet diverse user needs.
A larger model spends compute on queries a small model might answer correctly,
while a small model struggles with queries that need a stronger one.
Even among models of similar size, some are stronger in particular domains.
Routing sends each query to a suitable model in a pool, and cascading lets a
cheaper model answer first and escalates when its answer is not good enough
\citep{Moslem2026-RoutingSurvey}.

In the telecommunications domain, both accuracy and serving cost matter.
General-purpose models struggle with telecom standards
\citep{Maatouk2025-TeleQnA}, and operators are moving towards autonomous
networks, in which agents run workflows such as fault isolation and remediation
on models deployed on premises \citep{NVIDIA2026-TelcoReasoningModels}.
When models are served on premises, the cost of a model is the compute time it takes on fixed hardware,
under many concurrent requests, rather than a price per token.

Routers such as HybridLLM \citep{Ding2024-HybridLLM} and RouteLLM
\citep{Ong2025-RouteLLM} choose between a small and a large model before
generation, cascades such as FrugalGPT \citep{Chen2024-FrugalGPT} try models in
turn, and recent methods combine routing and cascading
\citep{Dekoninck2025-CascadeRouting,Shen2025-SATER}.
These methods price a model from an API price list, a token count, a recorded
run time or the share of queries sent to the smaller model.
They do not measure the time a request takes on the operator's own hardware under
concurrent load, which is the cost of serving on premises.

\begin{figure}[t]
\centering
\includegraphics[width=0.72\linewidth]{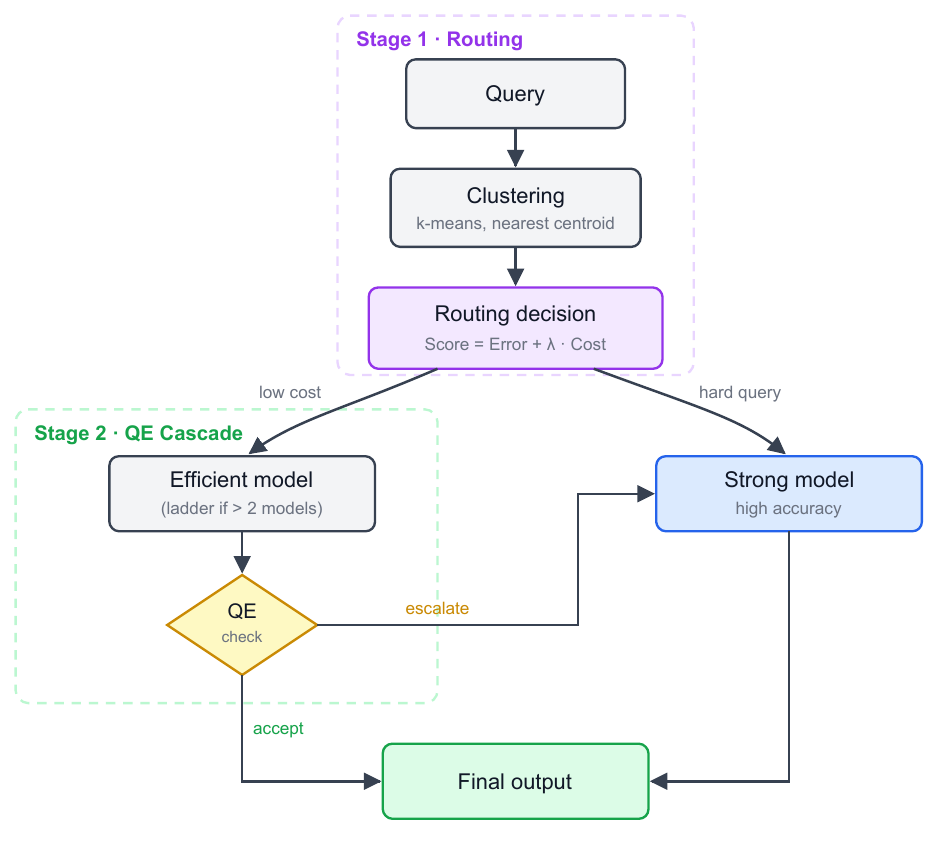}
\caption{\textbf{System overview.} Stage~1 assigns each query to a cluster and routes
it to the model chosen for that cluster under the cost budget. Stage~2 applies
where that model is cheaper than the strong model: a quality estimator reads the
answer, and escalates the query to the strong model when it rejects the answer.}
\label{fig:system}
\end{figure}

We present \textbf{CRE-Router} (Cluster, Route, Escalate; Figure~\ref{fig:system}),
which prices each model by the time it takes on the operator's GPUs under
concurrent load, per output token (TPOT) or per request (E2EL).
It combines routing and escalation because each catches what the other misses.
Routing decides before generation, so a query from a hard cluster goes straight
to a strong model and does not pay for a weak answer first.
Escalation decides after generation, so it can catch the errors that routing
cannot predict from the query alone.
Our contributions are the following.
\begin{itemize}
  \item \textbf{A two-stage method fitted on correctness labels alone.}
    Stage~1 clusters queries and routes each cluster to the model that best
    trades error against cost, under a budget on TPOT or E2EL. Stage~2 escalates 
    the queries whose answers a quality estimator rejects to the strong model.
  \item \textbf{An evaluation on the serving hardware.} We evaluate on two
    telecom benchmarks, TeleMath and TeleQnA, and on AIME, with pools of two to
    nine models served on one and two A100 GPUs per model. We compare against
    single models, and against FrugalGPT and HybridLLM, reimplemented on our
    pools for TeleMath and TeleQnA. Given the same cost per token,
    CRE-Router is significantly more accurate than FrugalGPT on TeleMath and
    TeleQnA. On all three benchmarks, CRE-Router is more accurate than every
    pool model that costs less per token, and comparable in accuracy to the
    strong model it escalates to at a lower cost per token.
  \item \textbf{Three findings about cost.} The cost term an operator chooses,
    TPOT or E2EL, decides which models serve the workload, because TPOT does
    not see how long the answers are. Conditioning cost on the cluster, rather
    than on the model as a whole, shortens the time per request by 20 to 22\%
    on TeleMath, at comparable accuracy. Hardware and load shift models' costs
    unevenly; at the same budget, one and two A100s select different routings
    on all three benchmarks, so a routing should be fitted again when the
    serving configuration changes.
  \item \textbf{Open-source release.} To facilitate reproducing our results and
    fitting CRE-Router to other model pools and hardware, we release the code,
    configurations, data and model outputs at\newline \coderepo
\end{itemize}

\section{Related Work}
\label{sec:related}

\paragraph{LLM routing.}
While model routing makes a single decision to map the query to one model,
model cascading operates sequentially, escalating to stronger models when
the initial response is insufficient \citep{Moslem2026-RoutingSurvey}.
\emph{HybridLLM} \citep{Ding2024-HybridLLM} trains a router that predicts how
much quality a query would lose on a small model rather than a large one, and
sends it to the small model when that loss is below a threshold that trades
quality for cost.
\emph{RouteLLM} \citep{Ong2025-RouteLLM} learns to route between a strong and a weak
model from human preference data; its routers include matrix factorisation and causal LLM
classifiers trained on Chatbot Arena data \citep{Chiang2024-ChatbotArena}.
\emph{IRT-Router} \citep{Song2025-IRT-Router} applies item response theory to model
query difficulty and LLM ability jointly, combining predicted correctness
probability with cost into a routing score for interpretable decisions.
\emph{Firewall routing} \citep{Peng2025-FirewallRoutingLLMs} routes between a
strong and a weak model, and sends to the weak model the long-tail queries on
which the strong model does no better, rather than spending the strong model on
them.
\emph{UniRoute} \citep{Jitkrittum2026-UniRoute} addresses dynamic routing where
previously unseen LLMs become available at test time, representing each model
as a feature vector of its prediction errors on a set of representative
prompts, such as its error in each of $K$ clusters of prompts.
Like \emph{UniRoute}, CRE-Router admits a new model without retraining the router;
only the new model answers the training queries, and the routing is fitted
again from the stored labels and costs.
Unlike these routers, CRE-Router is designed for self-hosting, where an operator
serves open-weight models on its own GPUs.
It measures each model's cost on that hardware under concurrent load, and can
condition that cost on the query cluster.
Where \emph{UniRoute} sweeps a cost penalty $\lambda$ to trace a trade-off curve
against API price or parameter count, CRE-Router selects the $\lambda^*$ that
meets an explicit budget on TPOT or E2EL, and it adds a second stage that
accepts or rejects each answer with a quality estimator and escalates the
queries whose answers it rejects.

\paragraph{LLM cascades.}
\emph{FrugalGPT} \citep{Chen2024-FrugalGPT} introduced cascade-based LLM serving
that queries a sequence of models adaptively, stopping when a quality
threshold is reached.
\emph{AutoMix} \citep{Aggarwal2024-AutoMix} uses self-verification to decide escalation.
Stage~1 of CRE-Router sends hard queries straight to the strong model under an
explicit cost budget, so they never pay for a small model's answer first.
Stage~2 escalates the queries whose answers a lightweight classifier, trained
on task-correctness labels, rejects.

\paragraph{Combining routing and cascading.}
Real-world multi-LLM routing systems are often compositional, integrating
several routing paradigms under operational constraints
\citep{Moslem2026-RoutingSurvey}.
\emph{Cascade-Routing} \citep{Dekoninck2025-CascadeRouting} unifies routing and
cascading.
At each step it chooses which model to run next, or whether to stop, from
per-query estimates of quality and cost under a cost budget, and it proves this
strategy optimal for the given estimates.
Its experiments price a model by API prices per token, tokens weighted by
model size, or
the completion times recorded for SWE-Bench agents \citep{Jimenez2024-SWEbench}.
\emph{SATER} \citep{Shen2025-SATER} fine-tunes the small model to give short answers
and to reject queries it cannot answer, which serves either routing before
generation or a cascade.
It uses API prices and takes output length as a proxy for latency, to remove
the effect of hardware.
Both thus price a model from a price list, a count or a recorded time, whereas our framework,
CRE-Router,
measures cost on the hardware that serves each model under concurrent load,
where a model's cost depends on the hardware and on the load
(Section~\ref{sec:res-hardware}).
Table~\ref{tab:methods} compares CRE-Router with the methods closest to it.

\begin{table}[ht]
\centering
\caption{\textbf{Routing and cascading methods at a glance.} When each method decides, how many models it chooses among, what it is trained on, and how its experiments price a model. We reimplement HybridLLM and FrugalGPT on our pool and data (Section~\ref{sec:baselines}). Only CRE-Router measures cost on the hardware that serves the models.}
\label{tab:methods}
\small
\begin{adjustbox}{max width=\textwidth}
\begin{tabular}{lllll}
\toprule
\textbf{Method} & \textbf{Decides} & \textbf{Models} & \textbf{Trained on} & \textbf{Cost measured as} \\
\midrule
HybridLLM \citep{Ding2024-HybridLLM} & before generation & 2 & quality gap (BARTScore; \citealp{Yuan2021-BARTScore}) & share of queries to the small model \\
RouteLLM \citep{Ong2025-RouteLLM} & before generation & 2 & human preferences & share of calls to the strong model, API prices \\
UniRoute \citep{Jitkrittum2026-UniRoute} & before generation & any number & answer correctness, per cluster & API prices, parameter count \\
FrugalGPT \citep{Chen2024-FrugalGPT} & after generation & any number, in lists of 3 & answer correctness & API prices \\
Cascade-Routing \citep{Dekoninck2025-CascadeRouting} & before and after & any number & per-query quality estimators & API prices, weighted tokens, recorded run time \\
SATER \citep{Shen2025-SATER} & before or after & 2 & short answers and rejections & API prices, output length for latency \\
\midrule
CRE-Router (ours) & before and after & any number & answer correctness & TPOT and E2EL measured on the serving hardware \\
\bottomrule
\end{tabular}
\end{adjustbox}
\end{table}

\paragraph{Adaptive inference and quality estimation.}
Estimating output quality without a reference signal enables adaptive decisions
at inference time.
\emph{Self-REF} \citep{Chuang2025-Self-REF} trains LLMs to emit confidence tokens
that trigger escalation, while \citet{Chuang2025-ConfidentSeekStronger}
benchmark uncertainty-driven routing from on-device SLMs to stronger LLMs.
\emph{CP-Router} \citep{Su2025-CP-Router} applies conformal prediction to route
between standard LLMs and large reasoning models based on output uncertainty.
\citet{Farinhas2025-TranslateSmart} propose quality-aware deferral in a
cascaded translation system, escalating outputs when a quality estimator
predicts failure; our Stage~2 applies the same deferral principle to
mathematics and telecom question answering, supervised by task correctness.
Adaptive thinking-length control
\citep{Zhang2025-AdaptThink,Zhang2025-ContinueThinking} adjusts compute per
query at the level of reasoning steps rather than model selection.
Speculative decoding \citep{Leviathan2023-SpeculativeDecoding}
accelerates generation within a single model, while our framework routes and escalates across models.

\paragraph{LLMs for telecommunications.}
\emph{TeleQnA} shows that general-purpose models struggle with questions on telecom
standards \citep{Maatouk2025-TeleQnA}, and \emph{TeleMath} that they struggle with
telecom mathematics, where models designed for reasoning do best
\citep{Colle2025-TeleMath}.
\emph{TeleCom-Bench} evaluates models on workflows taken from live network agents,
and finds that they handle the language of these tasks far better than their
procedures \citep{Xiao2026-TeleCom-Bench}.
We evaluate on two of these benchmarks, \emph{TeleQnA} and \emph{TeleMath}.

\section{Method: Cluster, Route, Escalate}
\label{sec:method}

\label{sec:system}
CRE-Router serves a pool of models under a cost budget the operator sets, and
chooses the most accurate routing whose cost stays within that budget.
The system comprises two stages (Figure~\ref{fig:system}).
Stage~1 (Section~\ref{sec:routing}) assigns an incoming query to a cluster and
sends it to the model that cluster is routed to.
We call the most accurate model a routing uses its strong model, and any model
cheaper than it an efficient model.
In clusters routed to an efficient model, Stage~2
(Section~\ref{sec:qe}) lets a quality estimator accept or reject each answer,
and escalates the queries whose answers it rejects to the strong model.
Section~\ref{sec:measured-cost} describes how each model's cost is measured on
the hardware that serves it.

The two stages have different roles.
Stage~1 decides before generation, for a whole cluster at once.
Where an efficient model usually answers wrongly, it sends the cluster straight
to a stronger model, so those queries never pay for an answer that is then
discarded, as they would in a cascade.
Stage~2 decides after generation, one query at a time.
It catches the wrong answers that remain in clusters routed to an efficient
model, which a decision per cluster cannot see.
Both stages are fitted offline, from the task-correctness labels of training
queries, which standard benchmark evaluation already provides.

\subsection{Stage 1: Clustering-Based Routing}
\label{sec:routing}

Routing maps each cluster of queries to one model.
It is fitted offline in five steps, described in turn below: cluster the
training queries, score each model on each cluster by its error and cost, prune
dominated models, sweep the cost penalty to find every possible routing, and
select the routing the budget allows.

\paragraph{Clustering.}
We adopt the query-embedding clustering of \citet{Jitkrittum2026-UniRoute}.
Training queries are encoded with \textit{all-MiniLM-L6-v2}
\citep{Wang2020-MiniLM,Reimers2019-SentenceTransformers} and clustered with $k$-means.
By default the number of clusters is chosen by the mean Silhouette score
\citep{Rousseeuw1987-Silhouettes} over $k \in [2, 9]$; it can also be fixed,
as we do on one benchmark (Section~\ref{sec:datasets}).
Centroids are fixed on the training data, and at inference a new query is
embedded once and assigned to its nearest centroid.

\paragraph{Routing score.}
Each cluster is routed to the model with the lowest score, which weighs the
model's error on that cluster against its serving cost measured on the
operator's hardware.
Let $\mathcal{M}$ be the model pool and $\mathcal{C}$ the clusters.
For each model $m$ and cluster $c$, $\mathrm{Error}(m, c)$ is the error rate on
the training queries of $c$, and $\mathrm{Cost}(m, c)$ is the mean serving cost
of those queries.
The score of model $m$ on cluster $c$ is
\begin{equation}
  \mathrm{Score}(m, c) = \mathrm{Error}(m, c) + \lambda \cdot
  \mathrm{Cost}_\mathrm{norm}(m, c),
  \label{eq:score}
\end{equation}
where $\lambda \geq 0$ sets how much accuracy the operator will trade for cost.
Models whose error lies within a tolerance $\epsilon$ of the minimiser's are
treated as tied, and the cheapest of them is chosen.
We set $\epsilon = 0.001$.
The tolerance can only move a choice towards a cheaper model and introduces no
new routing boundary; on the pools reported here it changes no routing.

\paragraph{Cost terms.}
$\mathrm{Cost}(m, c)$ is one of two measurements on the training queries,
as they were served (Section~\ref{sec:measured-cost}).
\emph{Time per output token} (TPOT) is the decoding time per generated token,
as reported by the serving stack.
\emph{End-to-end latency} (E2EL) is the time from sending a request to receiving
the complete answer.
TPOT ignores how long an answer is, while E2EL grows with it.
The two metrics can therefore rank models differently when models differ in
output length, for example when one reasons at length and another answers
directly.
We fit a separate routing for each metric and report both.
Streaming deployments also care about the time to first token (TTFT), which
E2EL includes.
On TeleMath, TTFT is under 2\% of E2EL for every model and cluster.
For a reasoning model, the first token opens the reasoning trace rather than
the answer, so TTFT does not show when the answer begins.

\paragraph{Cost conditioning.}
Equation~\ref{eq:score} conditions error on the cluster.
Cost can be conditioned in two ways.
Under \emph{per-model} cost, each model receives one cost, the mean over
clusters weighted by cluster size, which is the mean cost per training query,
and is normalised over the pool,
\begin{equation}
  \overline{\mathrm{Cost}}(m) = \frac{\sum_{c} n_c \, \mathrm{Cost}(m, c)}{\sum_{c} n_c},
  \qquad
  \mathrm{Cost}_\mathrm{norm}(m, c) =
  \frac{\overline{\mathrm{Cost}}(m) - \min_{m'} \overline{\mathrm{Cost}}(m')}
       {\max_{m'} \overline{\mathrm{Cost}}(m') - \min_{m'} \overline{\mathrm{Cost}}(m')},
  \label{eq:cost}
\end{equation}
where $n_c$ is the number of training queries in $c$.
The cheapest model in the pool thus has a normalised cost of 0, the most
expensive 1, and every other model falls in between, with the same value in
every cluster.
Under \emph{per-cluster} cost, each pair is priced on its own measurement, and
the min-max range is taken over all pairs, so that $\lambda$ keeps one meaning
across clusters,
\begin{equation}
  \mathrm{Cost}_\mathrm{norm}(m, c) =
  \frac{\mathrm{Cost}(m, c) - \min_{m', c'} \mathrm{Cost}(m', c')}
       {\max_{m', c'} \mathrm{Cost}(m', c') - \min_{m', c'} \mathrm{Cost}(m', c')}.
  \label{eq:cost-cluster}
\end{equation}
The methods we compare with price each model with a single cost, as
Equation~\ref{eq:cost} does; per-cluster cost is available to us because clustering happens offline, so the cost of serving each
cluster can be measured before any query arrives.
A model that is cheap on short queries and expensive on long ones is then priced
as such in each cluster rather than averaged across them.

\paragraph{Pareto pruning.}
A model is dominated if another is no more expensive and no less accurate in
every cluster, and strictly better in at least one of these.
The same tolerance $\epsilon$ applies to both comparisons.
Under per-cluster cost, ``no more expensive'' must hold in every cluster, which
is the stronger requirement, so tightening the rule can only enlarge the set of
surviving models.
A dominated model is never selected by Equation~\ref{eq:score} for any
$\lambda$, so it is removed before the sweep.

\paragraph{Routing regions.}
Sweeping $\lambda$ upwards from zero changes the assignment of a cluster only
at finitely many points, so the sweep partitions $\lambda$ into regions, each
with one fixed routing.
A boundary can only occur where two models tie in a cluster, at the $\lambda$
equal to their difference in error divided by their difference in normalised
cost.
The sweep therefore computes every such point for every pair of models and every
cluster, sorts them, and evaluates Equation~\ref{eq:score} once between each
pair of consecutive points, merging neighbouring intervals that give the same
routing.
As the boundaries are computed rather than searched for, the sweep is
exhaustive; it finds every region, which a grid search over $\lambda$ could
miss, and places each boundary exactly.
On TeleMath, the nine-model pool and four clusters give 11 regions under each
cost term on one A100.
Under per-cluster cost, normalised costs differ from cluster to cluster, so the
cost term can move a boundary and change the order in which clusters move to a
cheaper model.

For a pool of two models under per-model cost, such as our AIME pool
(Figure~\ref{fig:routing-regions}), the normalised costs are only 0 and 1, and
the boundary of cluster $c$ has a closed form,
\begin{equation}
  \lambda_c = \mathrm{Error}(m_\mathrm{eff}, c) - \mathrm{Error}(m_\mathrm{strong}, c),
  \label{eq:crossover}
\end{equation}
the efficient model's error penalty in that cluster.
Below $\lambda_c$ the cluster goes to the strong model, and above it to the
efficient one, so $N$ clusters give at most $N + 1$ regions.

\begin{figure}[t]
\centering
\begin{tikzpicture}[font=\small]
\draw[->, thick] (0,0) -- (11.0,0) node[right] {$\lambda$};
\fill[red!10] (0.0,0) rectangle (2.6,2.2);
\draw[thick] (0.0,0.12) -- (0.0,-0.12);
\node[below] at (0.0,-0.15) {$0$};
\node[align=center] at (1.3,1.1) {\textbf{R1}\\[2pt]\textcolor{red!70!black}{All Q30}};
\fill[orange!15] (2.6,0) rectangle (5.2,2.2);
\draw[thick] (2.6,0.12) -- (2.6,-0.12);
\node[below] at (2.6,-0.15) {$0.061$};
\node[align=center] at (3.9,1.1) {\textbf{R2}\\[2pt]C0:\,\textcolor{red!70!black}{Q30}\\C1:\,\textcolor{blue!70!black}{V}\\C2:\,\textcolor{red!70!black}{Q30}};
\fill[yellow!25] (5.2,0) rectangle (7.8,2.2);
\draw[thick] (5.2,0.12) -- (5.2,-0.12);
\node[below] at (5.2,-0.15) {$0.074$};
\node[align=center] at (6.5,1.1) {\textbf{R3}\\[2pt]C0:\,\textcolor{blue!70!black}{V}\\C1:\,\textcolor{blue!70!black}{V}\\C2:\,\textcolor{red!70!black}{Q30}};
\fill[blue!10] (7.8,0) rectangle (10.4,2.2);
\draw[thick] (7.8,0.12) -- (7.8,-0.12);
\node[below] at (7.8,-0.15) {$0.103$};
\node[align=center] at (9.1,1.1) {\textbf{R4}\\[2pt]\textcolor{blue!70!black}{All V}};
\end{tikzpicture}
\caption{\textbf{Routing regions on AIME.} Routing regions for the AIME two-model pool on one A100. As the cost penalty $\lambda$ increases, clusters move from Qwen3-30B-Thinking (Q30) to VibeThinker-1.5B (V) at the boundaries shown, in the order C1, C0, C2, the order of the error gap between the two models in each cluster. The regions are drawn at equal width.}
\label{fig:routing-regions}
\end{figure}

\paragraph{Budget-driven selection.}
The operator states a budget $B$ on system cost, in TPOT or E2EL.
Among the routing regions, we select the one with the highest training
accuracy whose system cost is within the budget,
\begin{equation}
  \lambda^* \in \argmax_{\mathrm{Cost}(\lambda) \leq B} \mathrm{Acc}(\lambda),
  \label{eq:lambda}
\end{equation}
where $\mathrm{Acc}(\lambda)$ and $\mathrm{Cost}(\lambda)$ are the training
accuracy and the size-weighted system cost of the routing in that region, and
ties go to the cheaper routing.
If no region meets $B$, the budget is infeasible for the pool.
Any $\lambda$ inside the selected region gives the same routing, so $B$ rather
than $\lambda$ is the parameter an operator sets.
The routing is fitted once on training data and applied unchanged to the test
set.
$B$ is a Stage~1 parameter.
Stage~2 adds escalations on top of it, and the reported costs include them.

\subsection{Stage 2: Escalation by a Quality Estimator}
\label{sec:qe}

Escalation catches the wrong answers that routing leaves behind.
As Stage~1 routes whole clusters, a cluster sent to an efficient model still
contains queries that model gets wrong.
Each routing escalates to one strong model, the most accurate model the routing
uses, by accuracy on the training queries.
Stage~2 gates only the clusters whose Stage~1 model is cheaper than the strong
model, under the cost term the routing was fitted on, since escalation only
makes sense from a cheaper model to a more expensive one.
Both follow from the fitted routing, and each gated model needs its own quality
estimator.
In a gated cluster, the estimator reads each answer and decides whether to
return it or to escalate the query to the strong model.

\paragraph{Quality estimator.}
\label{sec:qe-classifier}
We fine-tune ModernBERT-base \citep{Warner2025-ModernBERT} as a binary
classifier over the query, the efficient model's answer and the answer's length
in tokens.
At inference it runs once on each answer in a gated cluster.
Its labels are the task-correctness of the efficient model's answers on the
training queries, the same labels Stage~1 uses, so Stage~2 needs no further
annotation.
An answer is escalated when the estimated probability of acceptance is below a
threshold $\tau$; we use $\tau = 0.5$, the classifier's own decision.
Inputs, training data and hyperparameters for each benchmark are given in
Appendix~\ref{sec:appendix-qe-classifiers}.

\paragraph{What an escalation costs.}
\label{sec:qe-integration}
An escalated query is served twice, and the user waits for both passes.
Its E2EL is the sum of the two passes.
Its TPOT charges the decoding time of both passes to the tokens the user
receives, which are the strong model's,
\begin{equation}
  \mathrm{TPOT}_\mathrm{esc} =
  \frac{\mathrm{TPOT}_\mathrm{eff}\,(\ell_\mathrm{eff} - 1) + \mathrm{TPOT}_\mathrm{strong}\,(\ell_\mathrm{strong} - 1)}
       {\ell_\mathrm{strong} - 1},
  \label{eq:tpot-esc}
\end{equation}
where $\ell$ is the number of output tokens of each pass, and $\mathrm{TPOT}\,(\ell - 1)$ is
the decoding time of a pass under the serving stack's definition.
The discarded first answer therefore costs time and contributes no tokens.

\section{Experimental Setup}
\label{sec:setup}

\subsection{Benchmarks}
\label{sec:datasets}

We evaluate CRE-Router on three benchmarks with verifiable answers, chosen to differ in
domain, in answer length and in how much reasoning they reward.

\paragraph{TeleMath} \citep{Colle2025-TeleMath} contains 500 telecommunications mathematics problems
with numerical answers.
It ships no training split, so we draw one, stratified by category, giving 299
training and 201 test problems.
An answer is correct when it is within 1\% of the gold value.
TeleMath is our main benchmark.
Its pool is the largest, and it is the only benchmark on which both cost terms
and both cost-conditioning rules are compared; the prior methods are compared on
TeleMath under both cost terms and on TeleQnA under TPOT.

\paragraph{TeleQnA} \citep{Maatouk2025-TeleQnA} is a multiple-choice benchmark of
telecommunications knowledge.
We split it at random (seed 0) into 9,000 training and 1,000 test questions,
and release the split with our code.
An answer is correct when it selects the gold option.

\paragraph{AIME} (American Invitational Mathematics Examination) asks for an integer answer.
We train on the 921 problems of 1983 to 2023 and test on the 90 problems of
2024, 2025 and 2026.

\paragraph{Clusters.}
On TeleQnA and AIME the Silhouette score selects two and three clusters,
respectively.
On TeleMath we fix $k = 4$ rather than take the Silhouette choice of two, so
that a nine-model pool has more room to assign different models to different
parts of the workload.
The value is a design choice, not a fitted one.

\subsection{Model Pools}
\label{sec:pools}

Each benchmark has its own pool of open-weight models, served locally (Table~\ref{tab:pools}).
The AIME pool pairs VibeThinker-1.5B \citep{Xu2025-VibeThinker}, a small efficient model
trained for mathematics, with Qwen3-30B-A3B-Thinking (FP8) \citep{Yang2025-Qwen3}.
The TeleQnA pool holds Qwen3-4B-Instruct and three Gemma~4 models
\citep{ElAbd2026-Gemma4}, E2B, E4B and the 26B mixture of experts, all answering
without extended reasoning.
The TeleMath pool holds nine models from the Qwen~3 and Gemma~4 families, with
and without reasoning.
It includes Qwen3-30B-A3B-Thinking (FP8), so that the pool spans a wide range of
both accuracy and output length.
We chose the size of each pool, while Pareto pruning decides which models within 
a pool the router may use (Section~\ref{sec:routing}).

\begin{table}[t]
\centering
\caption{\textbf{Model pools.} Each benchmark's router chooses among these models, after Pareto pruning. Each cell shows whether the model reasons, writing an extended chain of thought before answering, when served in that pool. A cell reading \emph{on, off} means the same model is served both ways, as two pool members, and an empty cell means it is not in that pool. The pool size is given with each benchmark.}
\label{tab:pools}
\small
\begin{tabular}{lccc}
\toprule
\textbf{Hugging Face model} & \textbf{TeleMath (9)} & \textbf{TeleQnA (4)} & \textbf{AIME (2)} \\
\midrule
\texttt{google/gemma-4-26B-A4B-it} & on, off & off &  \\
\texttt{google/gemma-4-E2B-it} & on, off & off &  \\
\texttt{google/gemma-4-E4B-it} & on, off & off &  \\
\texttt{Qwen/Qwen3-30B-A3B-Thinking-2507-FP8} & on &  & on \\
\texttt{Qwen/Qwen3-4B-Instruct-2507} & off & off &  \\
\texttt{Qwen/Qwen3-4B-Thinking-2507} & on &  &  \\
\texttt{WeiboAI/VibeThinker-1.5B} &  &  & on \\
\bottomrule
\end{tabular}
\end{table}

\subsection{Serving}
\label{sec:serving}

Each model is served by its own vLLM~0.19.0 server \citep{Kwon2023-vLLM} on NVIDIA A100 SXM4
80\,GB GPUs, once with one GPU per model and once with two using tensor
parallelism.
Requests are sent at a concurrency of up to 32, and every configuration is run 5
times.
A run is one pass over a set of questions, sampling one answer to each.
Each of the efficient model's runs yields its own escalation set, served
separately, so pairing efficient and strong runs gives 25 measurements of a
system.
We report their mean, for every system including the baselines.
Sampling settings, output caps and the full serving configuration are given in
Appendix~\ref{sec:appendix-inference}.

\subsection{Measuring Cost on the Workloads Served}
\label{sec:measured-cost}

We call the set of queries one model serves in one measurement a
\emph{workload}.
Because a request's cost depends on the other requests served with it
(Section~\ref{sec:res-concurrency}), we price each request on the workload it
was served in.
Each model serves each test cluster as its own workload, so the Stage~1 of every
routing is assembled from these measurements.
An escalated query pays for both passes, the second priced on its escalation
set, which the strong model serves as a workload of its own on the same
hardware.
A single model serving the whole test set is measured on that set as one
workload.
The size of a workload also affects its cost, so a model can cost less per token
on a cluster than on the whole test set; Appendix~\ref{sec:appendix-basis}
prices the main TeleMath systems both ways as an ablation.

\subsection{Baselines}
\label{sec:baselines}

We compare CRE-Router with three kinds of baseline: single models serving every
question, two prior methods for routing and cascading, and our own system
without clustering.

\paragraph{Single models.}
Each pool member serves the whole test set alone.
Two models from outside the pools serve as references for a deployment that
serves one model, the dense Qwen3-8B and the Qwen3-Next-80B-A3B (FP8) mixture
of experts.

\paragraph{HybridLLM.}
HybridLLM \citep{Ding2024-HybridLLM} routes each query to one of two models
before generation, with a learned router and no escalation.
HybridLLM is defined for two models, so we give it the cheapest model in the pool
and the strong model of our routing under each cost term.
Its router can be trained on three kinds of label: deterministic, whether the
small model's answer is at least as good as the large model's; probabilistic,
the probability of that over sampled answers; and transformed, the same
probability with ``as good'' relaxed by a margin.
With correctness scores of 0 or 1, no margin below one can change a comparison,
so the transformed labels equal the probabilistic ones and only two schemes
remain.
We report the probabilistic router, which stands for both, as the main
HybridLLM result, and the deterministic one in
Appendix~\ref{sec:appendix-hybrid}.

\paragraph{FrugalGPT.}
Following its experiments, FrugalGPT \citep{Chen2024-FrugalGPT} searches every
ordered list of three models, fits a threshold per stage for each, and keeps the
most accurate cascade within the budget.
A scorer per model decides whether to return its answer.
On TeleMath it searches the five pool models our reported routings use, and on
TeleQnA the whole pool.

\paragraph{Fitting and pricing the prior methods.}
For a fair comparison, we reimplement both methods on TeleMath and TeleQnA with
our pools and training data, following each method's released code.
HybridLLM decides before generation, so it is given our Stage~1 budget.
FrugalGPT's budget bounds its whole cascade, escalations included, so we set it
to what our system spends at Stage~1+2, under either cost term, allowing it to
spend as much as our full system.
Each method's thresholds are fitted on the training data, as the most accurate
within its budget.
Both use ModernBERT-base, the encoder of our quality estimator, so that a
difference between methods is not a difference in model capacity.
The queries each sends to each model, and each set it escalates, are served as
their own workloads and priced the same way as ours.
While FrugalGPT measures cost by the prices of commercial APIs and HybridLLM by
the share of queries it sends to the small model, we measure both as TPOT or
E2EL on the hardware they are served on, so our results compare the methods as
they perform when self-hosted.

\paragraph{No clustering.}
To isolate what the clusters contribute, we fit the same pool with every
question in one cluster ($k = 1$), so Stage~1 selects a single model and
Stage~2 gates every query, escalating to the full system's strong model at the same
budget.

\subsection{Statistical Significance}
\label{sec:statistics}

For every comparison between two systems, we compute a 95\% confidence interval.
Each comparison has two sources of variance: the sampling of
the test questions from all questions a system could face, and the random
sampling of each model's answers \citep{Miller2024-ErrorBars}.
Our intervals account for both.
Following the same work, we compare two systems on paired differences per
question, so that each question's difficulty cancels out.
Each system answers every question once in each of its runs, so questions and
runs are crossed rather than nested, i.e., the results form a full grid of
questions by runs.
We therefore compute these intervals with the pigeonhole bootstrap
\citep{Owen2007-Pigeonhole} over 20,000 draws.
It resamples the questions once for both systems, and each system's runs
independently, since the two systems are served in separate runs.
Throughout the paper, a difference is referred to as ``significant'' when its
95\% interval excludes zero.
We report each comparison on its own and apply no correction for multiple
comparisons.
Appendix~\ref{sec:appendix-comparisons} lists every comparison the text makes,
with its interval.

\section{Results}
\label{sec:results}

\begin{figure}[t]
\centering
\includegraphics[width=\linewidth]{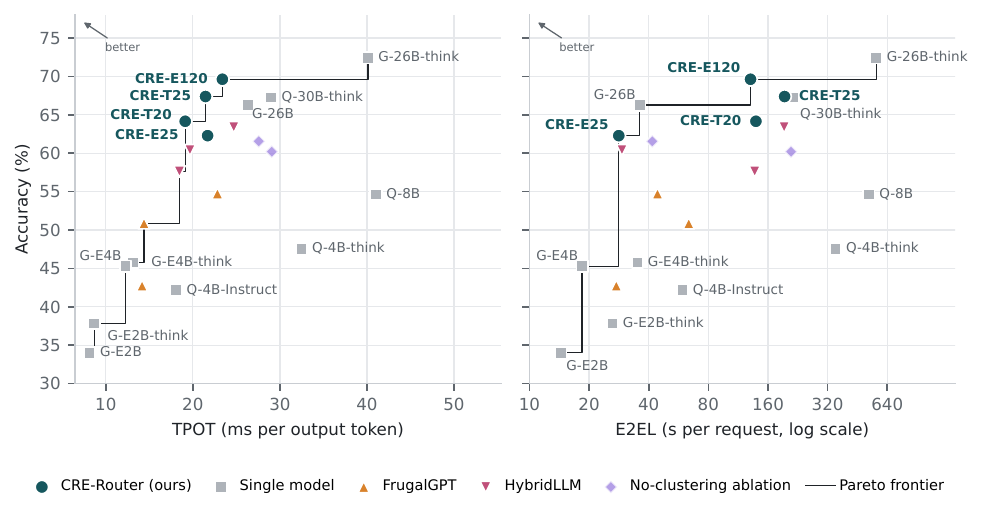}
\caption{\textbf{Pareto frontier of accuracy and cost on TeleMath.} Every system on the TeleMath test set on one A100, against its cost per output token (left) and per request (right, log scale): CRE-Router at Stage~1+2 under each reported budget, the pool and reference models serving every question alone, FrugalGPT and HybridLLM reimplemented on our pool and data, and the no-clustering ablation. Each CRE-Router point is labelled by the cost term and budget it was fitted under, T for TPOT and E for E2EL, so CRE-T20 is fitted under a TPOT budget of 20\,ms; the point sits at the cost it measured on the test set. Model labels abbreviate Gemma4 as G and Qwen3 as Q. Higher and further left is better; the line is the Pareto frontier, joining the systems that no other system beats on both accuracy and cost. A system identical to a single model is shown once, as that model, and the numbers are those of Tables~\ref{tab:telemath-main} and~\ref{tab:telemath-e2el}. On both cost terms, the CRE-Router routing fitted to that cost term lies on the frontier.}
\label{fig:telemath-pareto}
\end{figure}

We first compare CRE-Router with single models and with prior methods
(Sections~\ref{sec:res-single} and~\ref{sec:res-baselines}), and then measure
what each of its parts contributes (Section~\ref{sec:res-k1}).
We then report three findings about cost: the choice of cost term, conditioning
cost on the cluster, and the hardware and load the models are served under (Sections~\ref{sec:res-cost} to~\ref{sec:res-hardware}), and close with the overhead of the router itself (Section~\ref{sec:res-overhead}).
TeleMath, our main benchmark, carries the full set of comparisons on one A100.
Figure~\ref{fig:telemath-pareto} places every TeleMath system on both cost terms,
and Table~\ref{tab:telemath-main} gives the numbers at the two TPOT budgets.
AIME and TeleQnA follow in Sections~\ref{sec:res-single}
and~\ref{sec:res-baselines}, and Section~\ref{sec:res-hardware} reports all three
benchmarks on two A100s.

\begin{table}[t]
\centering
\caption{\textbf{Main results on TeleMath, TPOT budget.} CRE-Router against single models, two prior methods reimplemented on our pool and data, and its no-clustering ablation, on the TeleMath test set (201 questions, one A100, 32 concurrent requests). For CRE-Router, Stage 1 is routing alone and 1+2 adds escalation; HybridLLM, which decides before generation, is compared with Stage 1, and FrugalGPT, which decides after, with Stage 1+2. Accuracy, TPOT and E2EL are means over runs, each request priced on the workload it was served in. $B$ bounds the TPOT of Stage~1; FrugalGPT's budget bounds its whole cascade, so it is given the Stage~1+2 TPOT CRE-Router spends at each $B$, and HybridLLM is shown with its probabilistic router (Section~\ref{sec:baselines}). The two-model rows use HybridLLM's pair, Gemma4-E2B and Qwen3-30B-Thinking. At $B = 25$\,ms the two-model routing and the no-clustering ablation both send every question to Qwen3-30B-Thinking, so both equal that model serving alone. Within each budget, bold marks the most accurate system and underline the second. Within each budget CRE-Router is the most accurate system, and significantly more accurate than FrugalGPT.}
\label{tab:telemath-main}
\begin{tabular}{llrrr}
\toprule
\textbf{System} & \textbf{Stage} & \textbf{Acc.\ (\%)} & \textbf{TPOT (ms)} & \textbf{E2EL (s)} \\
\midrule
\multicolumn{5}{l}{\textit{Single models, whole test set}} \\[2pt]
Gemma4-26B-think & -- & 72.4 & 40.1 & 564.0 \\
Qwen3-30B-think & -- & 67.3 & 29.0 & 213.7 \\
Gemma4-26B & -- & 66.3 & 26.3 & 36.0 \\
Qwen3-8B & -- & 54.6 & 41.0 & 516.8 \\
Qwen3-4B-think & -- & 47.6 & 32.5 & 351.2 \\
Gemma4-E4B-think & -- & 45.8 & 13.1 & 35.1 \\
Gemma4-E4B & -- & 45.3 & 12.3 & 18.4 \\
Qwen3-4B-Instruct & -- & 42.2 & 18.1 & 59.2 \\
Gemma4-E2B-think & -- & 37.8 & 8.7 & 26.3 \\
Gemma4-E2B & -- & 34.0 & 8.1 & 14.4 \\
\midrule
\multicolumn{5}{l}{\textit{Budget $B = 20$\,ms}} \\[2pt]
HybridLLM (probabilistic) & -- & 56.8 & 19.3 & 149.2 \\
FrugalGPT & -- & 50.9 & 14.4 & 63.7 \\
\cmidrule{1-5}
No-clustering ablation ($k = 1$) & 1+2 & 60.2 & 29.1 & 209.2 \\
\arrayrulecolor{black!30}\cmidrule{1-5}\arrayrulecolor{black}
\multirow{2}{*}{\textbf{CRE-Router}, two models} & 1 & 62.3 & 18.8 & 136.2 \\
 & \textbf{1+2} & \textbf{65.0} & 19.8 & 148.0 \\
\multirow{2}{*}{\textbf{CRE-Router}, nine models (ours)} & 1 & 58.9 & 17.3 & 117.7 \\
 & \textbf{1+2} & \underline{64.2} & 19.1 & 139.1 \\
\midrule
\multicolumn{5}{l}{\textit{Budget $B = 25$\,ms}} \\[2pt]
HybridLLM (probabilistic) & -- & 63.4 & 24.7 & 193.1 \\
FrugalGPT & -- & 54.8 & 22.8 & 44.3 \\
\cmidrule{1-5}
No-clustering ablation ($k = 1$) & 1+2 & \underline{67.3} & 29.0 & 213.7 \\
\arrayrulecolor{black!30}\cmidrule{1-5}\arrayrulecolor{black}
\multirow{2}{*}{\textbf{CRE-Router}, two models} & 1 & \underline{67.3} & 29.0 & 213.7 \\
 & \textbf{1+2} & \underline{67.3} & 29.0 & 213.7 \\
\multirow{2}{*}{\textbf{CRE-Router}, nine models (ours)} & 1 & 64.7 & 20.4 & 182.3 \\
 & \textbf{1+2} & \textbf{67.4} & 21.5 & 194.1 \\
\bottomrule
\end{tabular}
\end{table}

\subsection{Comparison with Single Models}
\label{sec:res-single}

A single model here is one model serving every test question alone: each member
of the pool, and the two reference models from outside it
(Section~\ref{sec:baselines}).

\paragraph{Against the pool.}
Table~\ref{tab:telemath-main} compares CRE-Router with every single model on
TeleMath, and Figure~\ref{fig:telemath-pareto} shows the accuracy--cost Pareto frontier
under each cost term.
At both budgets ($B = 20$ and 25\,ms), CRE-Router is at least ten points more accurate than every pool
model with a lower TPOT.
Against Qwen3-30B-Thinking, its strong model, CRE-Router is comparable in accuracy
at both budgets, with a 26\% lower TPOT at $B = 25$\,ms and a larger saving at
$B = 20$\,ms.

\paragraph{Against the two Gemma4-26B models (with and without reasoning).}
CRE-Router is comparable in accuracy to Gemma4-26B without reasoning at both
budgets, with a 19 to 27\% lower TPOT but an E2EL four to five times higher,
because that model's answers are short.
Which of the two is the better deployment therefore depends on the cost term,
which Section~\ref{sec:res-cost} takes up.
CRE-Router is less accurate than Gemma4-26B with reasoning, the most accurate
model in the pool, significantly so at $B = 20$\,ms, but has a 47 to 52\% lower
TPOT and a 66 to 75\% lower E2EL; no budget we consider admits that model.

\paragraph{Against reference models.}
Against the dense Qwen3-8B, our reference for a single-model deployment,
CRE-Router is significantly more accurate at both budgets, with a 48 to 53\%
lower TPOT.
Our second reference, Qwen3-Next-80B, needs two A100s and is compared in
Section~\ref{sec:res-hardware}.

\paragraph{On AIME and TeleQnA.}
The same pattern holds on the other two benchmarks (Table~\ref{tab:appendix-full}).
On AIME, on one A100 and on two, CRE-Router is significantly more accurate than
VibeThinker-1.5B, its efficient model, and comparable in accuracy to
Qwen3-30B-Thinking, its strong model, with a significantly lower TPOT and E2EL.
On TeleQnA, CRE-Router is significantly more accurate than each efficient model
and within two points of Gemma4-26B, its strong model, with a significantly lower
TPOT.

\subsection{Comparison with Baselines}
\label{sec:res-baselines}

We compare CRE-Router with FrugalGPT and HybridLLM on TeleMath under each cost term, and on TeleQnA under TPOT, each reimplemented on our pools and given the same budget (Section~\ref{sec:baselines}).

\paragraph{FrugalGPT.}
CRE-Router is significantly more accurate than FrugalGPT at both TPOT budgets, by
more than twelve points, although FrugalGPT is allowed to spend on its whole
cascade the TPOT that CRE-Router spends at Stage~1+2.
FrugalGPT's fitted cascades answer most questions with their first model, Gemma4-E4B with
reasoning, which is cheap but answers fewer than half the questions correctly, and
escalate few of them.
Fitted at every budget instead (Figure~\ref{fig:frugal-sweep}), FrugalGPT needs a
TPOT of about 26 and 29\,ms to match the accuracy CRE-Router reaches at 19.1 and
21.5\,ms for the full system.

\paragraph{HybridLLM.}
HybridLLM routes between two models before generation, so we compare it with
CRE-Router on the same two models.
We show its probabilistic router (Section~\ref{sec:baselines}).
CRE-Router's Stage~1 alone, the counterpart of HybridLLM, is more accurate at both budgets,
significantly so at $B = 20$\,ms.
With escalation, CRE-Router on the same two models is significantly more
accurate than HybridLLM at $B = 20$\,ms.
At $B = 25$\,ms the two-model routing sends every cluster to Qwen3-30B, so it is
that model serving alone, and its two rows in the table are the same.
HybridLLM is also limited to two models by construction, while CRE-Router fits the whole pool; at $B = 25$\,ms, CRE-Router's Stage~1 over the full pool has a significantly lower TPOT than HybridLLM, at comparable accuracy.

\begin{table}[t]
\centering
\caption{\textbf{Results on TeleMath, E2EL budget.} TeleMath, one A100, measured as in Table~\ref{tab:telemath-main}. FrugalGPT is given the E2EL CRE-Router spends at Stage~1+2, 28.2\,s, the rule Table~\ref{tab:telemath-main} applies to TPOT. The first block is every system at the E2EL budget $B = 25$\,s, and bold and underline mark its most and second most accurate. On HybridLLM's two models CRE-Router selects the same routing as on the full pool, so that row is not repeated. The second block holds CRE-Router at a looser budget, 120\,s, at which no prior method was fitted, and Gemma4-26B serving every question alone. Given a budget of 35\,s, FrugalGPT's cascade sends every question to Gemma4-26B and so becomes that model, so it is not shown separately.}
\label{tab:telemath-e2el}
\begin{tabular}{llrrr}
\toprule
\textbf{System} & \textbf{Stage} & \textbf{Acc.\ (\%)} & \textbf{TPOT (ms)} & \textbf{E2EL (s)} \\
\midrule
\multicolumn{5}{l}{\textit{Budget $B = 25$\,s}} \\
HybridLLM & -- & 60.4 & 19.7 & 29.3 \\
FrugalGPT & -- & 42.8 & 14.2 & 27.4 \\
No-clustering ablation ($k = 1$) & 1+2 & \underline{61.6} & 27.6 & 41.7 \\
\textbf{CRE-Router}, nine models (ours) & 1 & 60.0 & 19.3 & 25.8 \\
 & 1+2 & \textbf{62.3} & 21.7 & 28.2 \\
\midrule
\multicolumn{5}{l}{\textit{Looser budgets}} \\
CRE-Router (ours), $B = 120$\,s & 1 & 69.1 & 23.7 & 109.6 \\
 & 1+2 & 69.6 & 23.4 & 130.6 \\
\midrule
Gemma4-26B, whole test set & -- & 66.3 & 26.3 & 36.0 \\
\bottomrule
\end{tabular}
\end{table}

\paragraph{Comparisons at the E2EL budget.}
Table~\ref{tab:telemath-e2el} repeats this comparison at an E2EL budget of 25\,s.
At this budget, the E2EL of CRE-Router on the test set exceeds 25\,s, because $B$ bounds the Stage~1 cost fitted on training data, and the test workloads cost more.
CRE-Router is significantly more accurate than FrugalGPT, at a comparable E2EL, although FrugalGPT is allowed to spend the E2EL that CRE-Router spends at Stage~1+2.
CRE-Router is comparable to HybridLLM and the no-clustering ablation in accuracy
with a significantly lower E2EL than either.
CRE-Router is about four points less accurate than Gemma4-26B serving every question alone, with a significantly lower E2EL.
Fitted at every budget from 15 to 600\,s (Figure~\ref{fig:frugal-sweep}),
FrugalGPT stays below 45\% up to 30\,s and never exceeds 72\%.
Given a looser budget, FrugalGPT's cascade sends every question to Gemma4-26B and
so becomes that model, no longer a cascade.
On this benchmark and cost term, a single model is therefore a strong
alternative, and CRE-Router's contribution is to trade some of its accuracy for a lower E2EL.

\paragraph{On TeleQnA.}
Table~\ref{tab:teleqna-baselines} repeats the comparison on TeleQnA at the
reported budget.
CRE-Router is again significantly more accurate than FrugalGPT, whose cascade
answers most questions at its first model, Qwen3-4B-Instruct, at a comparable TPOT and E2EL.
At no TPOT budget from 10 to 32\,ms does FrugalGPT exceed 70.4\%
(Figure~\ref{fig:frugal-sweep}): its scorers, fitted on the training answers, put
its training accuracy above that of Gemma4-26B alone, so beyond 25\,ms its fit
no longer changes.
CRE-Router's Stage~1 over the full pool is comparable in accuracy to HybridLLM's probabilistic router, with a significantly lower TPOT and E2EL.
On HybridLLM's two models, CRE-Router's Stage~1 has a significantly lower TPOT and is less accurate, and escalation makes it significantly more accurate than
HybridLLM, at a higher TPOT.

\begin{table}[t]
\centering
\caption{\textbf{Prior methods on TeleQnA, TPOT budget.} CRE-Router against HybridLLM and FrugalGPT on the TeleQnA test set (1,000 questions, one A100, 32 concurrent requests), at a TPOT budget of $B = 20$\,ms, each row priced on the workloads it served. HybridLLM, which decides before generation, is compared with Stage~1, and FrugalGPT, which decides after, with Stage~1+2. FrugalGPT is given the Stage~1+2 TPOT CRE-Router spends at this budget, 24.1\,ms, and HybridLLM is shown with its probabilistic router. The two-model rows use HybridLLM's pair, Gemma4-E2B and Gemma4-26B. Bold marks the most accurate system and underline the second.}
\label{tab:teleqna-baselines}
\begin{tabular}{llrrr}
\toprule
\textbf{System} & \textbf{Stage} & \textbf{Acc.\ (\%)} & \textbf{TPOT (ms)} & \textbf{E2EL (s)} \\
\midrule
HybridLLM (probabilistic) & -- & 72.2 & 19.1 & 0.92 \\
FrugalGPT & -- & 70.1 & 24.9 & 1.16 \\
\cmidrule{1-5}
\multirow{2}{*}{\textbf{CRE-Router}, two models} & 1 & 69.9 & 17.0 & 0.78 \\
 & \textbf{1+2} & \underline{74.9} & 23.6 & 1.15 \\
\multirow{2}{*}{\textbf{CRE-Router}, four models (ours)} & 1 & 71.0 & 18.2 & 0.85 \\
 & \textbf{1+2} & \textbf{75.1} & 24.1 & 1.18 \\
\bottomrule
\end{tabular}
\end{table}

\subsection{Ablation Studies}
\label{sec:res-k1}

Each ablation removes one part of CRE-Router with everything else fixed: the
clusters, the escalation, the estimator's choice of which answers to escalate,
and the size of the pool.

\paragraph{Clustering.}
Removing the clusters, so that Stage~1 chooses one model for every question
and Stage~2 gates every query, isolates what the clustering is worth
(Table~\ref{tab:telemath-main}).
With everything else fixed, CRE-Router is comparable in accuracy to the no-clustering ablation, with a 26 to 34\% lower TPOT at both budgets.
The gain from clustering is in cost, not accuracy.
It lets Stage~1 send each part of the workload to a cheaper model where that
model is good enough, which a single choice for the whole workload cannot do.

\paragraph{Escalation.}
At both budgets in Table~\ref{tab:telemath-main}, Stage~2 raises accuracy over
Stage~1 of the same system significantly, while adding one to two milliseconds of TPOT.
The gain is larger at the tighter budget, where Stage~1 routes more clusters to
efficient models and so leaves more answers worth checking.
The budget bounds Stage~1, and Stage~2 adds to it, since each escalated query
also pays for the strong model's pass.
On TeleMath it adds up to 2.3\,ms of TPOT (up to 14\%) and 2 to 59\,s of E2EL
(6 to 34\%) across the reported routings, and on AIME at most 8\% of either.
Across all three benchmarks and both card counts, the increment can be estimated
from training data alone with an error of at most 7\% of the total E2EL, so it can
be anticipated before deployment and the budget set below the total an operator
can afford (Appendix~\ref{sec:appendix-stage2-cost}).

\begin{table}[t]
\centering
\caption{\textbf{Contribution of the estimator.} What the quality estimator contributes: accuracy (\%) when no answer is escalated, when the estimator chooses which to escalate, when the same number is escalated at random (expected), when the longest answers are escalated first, and when every answer is. On AIME and TeleQnA every question is gated; on TeleMath the deployed routing gates its two clusters at the reported threshold. The last column is the estimator's gain over random escalation with its 95\% paired interval. The estimator beats random escalation significantly on AIME and TeleQnA, and is comparable to it on TeleMath; on AIME, where accept probability tracks answer length closely, escalating the longest answers does as well.}
\label{tab:stage2}
\small
\begin{adjustbox}{max width=\textwidth}
\begin{tabular}{llrrrrrrl}
\toprule
\textbf{Benchmark} & \textbf{Gating} & \textbf{Gated} & \textbf{None} & \textbf{Random} & \textbf{Estimator} & \textbf{Longest} & \textbf{All} & \textbf{Estimator $-$ random} \\
\midrule
TeleMath & deployed routing, $B = 20$\,ms & 91 & 58.9 & 62.5 & 64.2 & -- & 69.1 & $+1.7$ [$-0.1$, $+3.5$] \\
TeleQnA & every question gated & 1{,}000 & 66.6 & 70.5 & 73.5 & -- & 76.8 & $+3.0$ [$+1.6$, $+4.4$] \\
AIME & every question gated & 90 & 74.0 & 77.0 & 83.1 & 84.1 & 85.8 & $+6.1$ [$+2.6$, $+10.2$] \\
\bottomrule
\end{tabular}
\end{adjustbox}
\end{table}

\paragraph{What the estimator selects.}
To separate what the estimator chooses from the benefit of escalating at all, we
compare it with escalating the same number of answers at random
(Table~\ref{tab:stage2}).
On AIME and TeleQnA, where every question is gated, the estimator is
significantly better than random; on TeleMath, in the deployed routing, it is
comparable to random escalation.
On AIME, escalating the longest answers first does as well as the estimator,
whose accept probability tracks answer length closely; on TeleQnA, whose answers
are short multiple-choice explanations, length carries almost no signal.

\paragraph{Pool size.}
Over its full pool, CRE-Router is comparable in accuracy to CRE-Router restricted
to the two models HybridLLM uses, with a lower TPOT at both budgets; at
$B = 25$\,ms the two-model routing can only send every question to Qwen3-30B,
which exceeds the budget (Appendix~\ref{sec:appendix-pool}).

\subsection{Effect of the Cost Term}
\label{sec:res-cost}

\begin{table}[t]
\centering
\caption{\textbf{Routings by cost term.} The routing CRE-Router selects on TeleMath under each cost term, one A100. Each budget is fitted on training costs and the routing is then measured on the test set, Stage~1+2. The TPOT budgets select Qwen3-30B with reasoning for two of the four clusters, whose long answers make every request slow; the E2EL budget selects Gemma4-26B without reasoning instead, at a similar cost per token and a fraction of the wait per request. A looser E2EL budget of 120\,s buys back Qwen3-30B for two clusters and escalates the other two to it. The last column counts the A100s a routing loads, one per model it uses; Stage~2 escalates to one of those models, so it needs no further card.}
\label{tab:telemath-routings}
\small
\begin{adjustbox}{max width=\textwidth}
\begin{tabular}{lllllrrrr}
\toprule
\textbf{Budget} & \textbf{C0} & \textbf{C1} & \textbf{C2} & \textbf{C3} & \textbf{Acc.\ (\%)} & \textbf{TPOT (ms)} & \textbf{E2EL (s)} & \textbf{A100s} \\
\midrule
TPOT 20\,ms & Qwen3-30B-think & Gemma4-E2B & Qwen3-30B-think & Gemma4-E4B-think & 64.2 & 19.1 & 139.1 & 3 \\
TPOT 25\,ms & Qwen3-30B-think & Gemma4-E2B & Qwen3-30B-think & Gemma4-26B-think & 67.4 & 21.5 & 194.1 & 3 \\
E2EL 25\,s & Gemma4-26B & Gemma4-E2B & Gemma4-26B & Gemma4-26B & 62.3 & 21.7 & 28.2 & 2 \\
E2EL 120\,s & Qwen3-30B-think & Qwen3-30B-think & Gemma4-26B & Gemma4-26B & 69.6 & 23.4 & 130.6 & 2 \\
\bottomrule
\end{tabular}
\end{adjustbox}
\end{table}

\begin{figure}[t]
\centering
\includegraphics[width=0.85\linewidth]{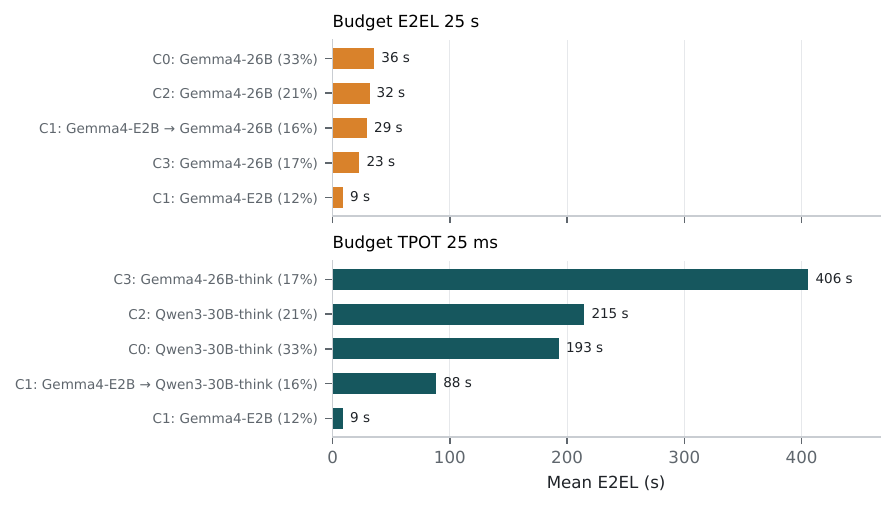}
\caption{\textbf{Time per request by cost term.} Every path a TeleMath request can take through CRE-Router at Stage~1+2 on one A100, under the E2EL budget of 25\,s (top) and the TPOT budget of 25\,ms (bottom): the cluster, the model that answers it (an arrow marks an escalation), the share of requests that take the path, and their mean E2EL. Under the TPOT budget, the more than 70\% of requests routed to reasoning models wait more than three minutes on average; under the E2EL budget no path averages more than 40 seconds.}
\label{fig:served-wait}
\end{figure}

\paragraph{The two cost terms select different routings.}
Table~\ref{tab:telemath-routings} shows the routing CRE-Router selects on
TeleMath under each cost term.
Under a TPOT budget it sends two of the four clusters to Qwen3-30B with
reasoning; under an E2EL budget it sends them to Gemma4-26B without reasoning.
The two kinds of routing have about the same TPOT, but the TPOT routings have an E2EL five to seven times higher, because TPOT does not
see how many tokens an answer takes.
A budget on TPOT therefore admits long reasoning that a budget on E2EL rules
out, and the operator's choice of cost term decides which models serve the
workload.
Figure~\ref{fig:served-wait} shows what this means for the requests themselves.

The divergence is a property of the pool rather than of one model in it.
Removing any single model from the nine still leaves the two cost terms
selecting different routings at both TPOT budgets.
The divergence also holds at every budget other than the ones we report, except
the loosest and the tightest, where both terms select the most accurate routing
and the cheapest one (Appendix~\ref{sec:appendix-routings}).
Gemma4-26B without reasoning, which serves all but two of the routings an E2EL
budget can select, appears in none of those a TPOT budget can select.
At a budget of 25\,s, the E2EL routing is less accurate than either TPOT
routing, in exchange for a lower E2EL.
A looser E2EL budget of 120\,s removes that loss
(Tables~\ref{tab:telemath-routings} and~\ref{tab:telemath-e2el}).
Its routing is comparable in accuracy to the TPOT routing at 25\,ms and has a significantly lower E2EL, at a 9\% higher TPOT.
Against Qwen3-30B-Thinking serving every question, it is comparable in accuracy
and has a significantly lower TPOT and E2EL.

\paragraph{Where a per-token budget costs time per request.}
At $B = 25$\,ms, the TPOT of CRE-Router is 26\% lower than that of Qwen3-30B-Thinking, yet its E2EL is about the same.
One cluster accounts for the difference.
The TPOT budget sends C3 to Gemma4-26B with reasoning, which has a low TPOT but writes long answers, so on that cluster a request takes more than twice as
long as with Qwen3-30B.
Requests in the other three clusters finish sooner, and the slow cluster offsets
most of that saving.

\subsection{Per-Cluster Cost Conditioning}
\label{sec:res-percluster}

\begin{table}[t]
\centering
\caption{\textbf{Per-cluster cost conditioning.} CRE-Router on TeleMath with per-model and per-cluster cost, one A100, Stage~1+2 on the test set. Both routings escalate the same cluster, C1, with the same estimator, so they differ in the cost rule alone. With per-cluster cost the accuracy is the same, each request waits 20 to 22\% less, and the cost per token is the same at $B = 25$\,ms and 0.7\,ms higher at $B = 20$\,ms, still within the budget. The per-cluster routing also sends C2 to Gemma4-26B, which is cheaper than Qwen3-30B in that cluster and could be escalated as well; doing so gives 65.2\%, 19.7\,ms and 123.9\,s at $B = 20$\,ms and 67.6\%, 21.3\,ms and 170.0\,s at $B = 25$\,ms.}
\label{tab:telemath-percluster}
\small
\begin{adjustbox}{max width=\textwidth}
\begin{tabular}{llllllrrr}
\toprule
\textbf{Budget} & \textbf{Cost} & \textbf{C0} & \textbf{C1} & \textbf{C2} & \textbf{C3} & \textbf{Acc.\ (\%)} & \textbf{TPOT (ms)} & \textbf{E2EL (s)} \\
\midrule
20\,ms & per-model & Qwen3-30B-think & Gemma4-E2B & Qwen3-30B-think & Gemma4-E4B-think & 64.2 & 19.1 & 139.1 \\
 & per-cluster & Qwen3-30B-think & Gemma4-E2B & Gemma4-26B & Qwen3-30B-think & 64.8 & 19.8 & 108.8 \\
\midrule
25\,ms & per-model & Qwen3-30B-think & Gemma4-E2B & Qwen3-30B-think & Gemma4-26B-think & 67.4 & 21.5 & 194.1 \\
 & per-cluster & Qwen3-30B-think & Gemma4-E2B & Gemma4-26B & Gemma4-26B-think & 67.2 & 21.4 & 154.9 \\
\bottomrule
\end{tabular}
\end{adjustbox}
\end{table}

Per-cluster cost (Section~\ref{sec:routing}) prices each model separately in each
cluster, rather than with one cost for the whole workload.
On TeleMath, at both TPOT budgets on one A100, it gives CRE-Router a 20 to 22\%
lower E2EL at comparable accuracy, with a TPOT that stays within the budget
(Table~\ref{tab:telemath-percluster}).
It sends C2 to Gemma4-26B rather than Qwen3-30B with reasoning, as Gemma4-26B is
the cheaper of the two on that cluster although not on average; at
$B = 20$\,ms it also moves C3 from Gemma4-E4B with reasoning to Qwen3-30B.
Both routings escalate the same cluster with the same estimator, so they differ
only in the cost rule.

Per-cluster cost can change a routing only when a model is cheaper than another
in one cluster and more expensive in another.
On AIME and TeleQnA no pair of models crosses in this way, so both rules select
the same routing.
On TeleMath five pairs cross on one A100 and two on two A100s, but only at the
TPOT budgets on one A100 does this change the routing.

\subsection{Hardware and Load}
\label{sec:res-hardware}

\begin{figure}[t]
\centering
\includegraphics[width=0.62\linewidth]{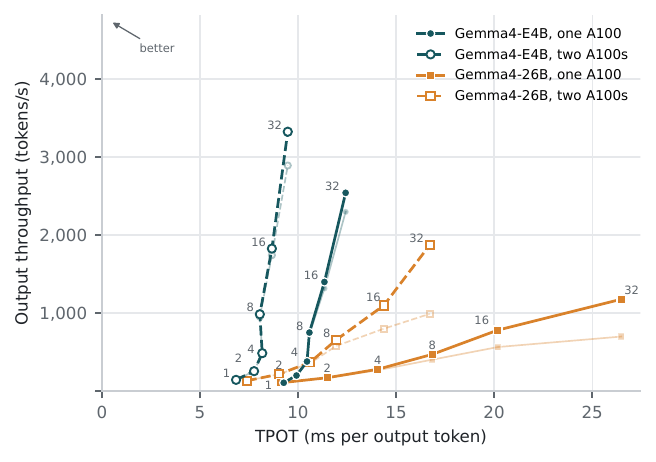}
\caption{\textbf{Throughput and cost under concurrent load.} Output throughput against TPOT for Gemma4-E4B and Gemma4-26B serving the TeleMath test set without reasoning, on one A100 (solid lines, filled markers) and two (dashed lines, hollow markers), as the number of concurrent requests doubles from 1 to 32 along each line; the number beside each point is its number of concurrent requests, and the four 1-request points, which sit together, are labelled once. Dark lines give the throughput while every request slot is busy; light lines give vLLM's figure over the whole run, which includes the end-of-run drain. Moving along a line, more requests in flight buy more throughput at a higher cost per token. The numbers are those of Table~\ref{tab:concurrency}.}
\label{fig:concurrency}
\end{figure}

The cost of a model depends on the hardware that serves it and on the load it
serves.
We vary both: Table~\ref{tab:hardware} reports CRE-Router and two reference
models on one and two A100s, and Figure~\ref{fig:concurrency} varies the number
of concurrent requests.

\begin{table}[t]
\centering
\caption{\textbf{Hardware sensitivity.} CRE-Router (Stage~1+2) and two reference models on one and two A100s, each model on two cards using tensor parallelism. Each card count is fitted on its own measured costs, so its budget $B$ (TPOT, ms) differs; Qwen3-Next-80B does not fit on one A100. Within each benchmark and card count, bold marks the most accurate system and underline the second. Every system in the table costs less per token on two cards. The reported routings differ between card counts only on TeleMath, as each card count is reported at its own budget.}
\label{tab:hardware}
\small
\begin{adjustbox}{max width=\textwidth}
\begin{tabular}{llrrrrrr}
\toprule
 & & \multicolumn{3}{c}{\textbf{1$\times$ A100}} & \multicolumn{3}{c}{\textbf{2$\times$ A100}} \\
\cmidrule(lr){3-5}\cmidrule(lr){6-8}
\textbf{Benchmark} & \textbf{System} & \textbf{Acc.\ (\%)} & \textbf{TPOT (ms)} & \textbf{E2EL (s)} & \textbf{Acc.\ (\%)} & \textbf{TPOT (ms)} & \textbf{E2EL (s)} \\
\midrule
TeleMath & Qwen3-8B & \underline{54.6} & 41.0 & 516.8 & 54.4 & 21.3 & 270.7 \\
 & Qwen3-Next-80B-A3B-Thinking & -- & -- & -- & \underline{69.7} & 18.2 & 142.2 \\
 & CRE-Router (ours), $B = 25$ / $20$\,ms & \textbf{67.4} & 21.5 & 194.1 & \textbf{73.3} & 19.0 & 232.4 \\
\midrule
TeleQnA & Qwen3-8B & \underline{73.4} & 15.8 & 14.1 & 73.3 & 9.7 & 8.6 \\
 & Qwen3-Next-80B-A3B-Instruct & -- & -- & -- & 72.7 & 36.4 & 2.0 \\
 & Qwen3-Next-80B-A3B-Thinking & -- & -- & -- & \textbf{76.7} & 17.6 & 24.6 \\
 & CRE-Router (ours), $B = 20$ / $15$\,ms & \textbf{75.1} & 24.1 & 1.2 & \underline{74.7} & 16.9 & 0.8 \\
\midrule
AIME & Qwen3-8B & \underline{70.4} & 42.0 & 711.0 & 69.8 & 23.1 & 404.8 \\
 & Qwen3-Next-80B-A3B-Thinking & -- & -- & -- & \textbf{88.0} & 19.0 & 357.5 \\
 & CRE-Router (ours), $B = 30$ / $20$\,ms & \textbf{85.1} & 20.1 & 362.7 & \underline{87.0} & 13.5 & 246.1 \\
\bottomrule
\end{tabular}
\end{adjustbox}
\end{table}

\paragraph{A second card changes models' costs unevenly.}
Serving a model on two A100s with tensor parallelism lowers its TPOT by very different amounts.
On the TeleMath training set, a second card divides a model's TPOT by between 1.0 and 1.7, depending on the model (Table~\ref{tab:appendix-hw-speed}).
Since Equation~\ref{eq:cost} normalises cost over the pool, an uneven change
reorders the trade-off between models, and the same budget no longer names the
same system.
We fit each card count on its own measured costs.

\paragraph{The selected routing follows the hardware.}
At the same budget, one and two A100s select different routings on all three
benchmarks (Table~\ref{tab:appendix-hw-samebudget}).
On AIME and TeleQnA, a budget at which one card mixes models sends every
cluster to the strong model on two, so each card count is reported at a budget
at which its routing still mixes models.
A routing fitted on one serving configuration is therefore not guaranteed to
remain the right one on another, and a deployment that changes its hardware
should fit again; only the costs need measuring again, as the correctness labels stay the same.
On TeleMath, the two-card routing lies on the Pareto frontier of both cost
terms (Appendix~\ref{sec:appendix-two-card}); it is the only routing we fitted on
two cards, and no prior method was run there.

\paragraph{Against reference models.}
On two A100s, Table~\ref{tab:hardware} also compares CRE-Router with
Qwen3-Next-80B, a mixture of experts with 80 billion parameters that does not fit
on one; the largest model in CRE-Router's pools has 30 billion.
CRE-Router is comparable in accuracy to Qwen3-Next-80B with reasoning on AIME and
TeleMath, and to both of its versions on TeleQnA.
It has a lower TPOT and E2EL on AIME and TeleQnA, but on TeleMath its TPOT is 4\%
higher and its E2EL 63\% higher.
Against the dense Qwen3-8B, CRE-Router is significantly more accurate on AIME and
TeleMath on both card counts, and comparable in accuracy on TeleQnA.

\paragraph{Load changes the ratio between models.}
\label{sec:res-concurrency}
Every cost in this paper is measured at a concurrency of up to 32.
Figure~\ref{fig:concurrency} varies that load for two models of similar active
size, a dense model and a mixture of experts, to show how much the choice of load matters;
Appendix~\ref{sec:appendix-concurrency} gives the values and the throughput.
From one request to 32, the dense model's TPOT rises by 34 to 39\%, while the mixture of experts' nearly triples on one A100 and more than
doubles on two.
On one A100 the two have about the same TPOT for a single request, and at 32 the mixture of experts has about twice the TPOT.
Load therefore changes not only every model's cost but the ratio between
models, which is what Equation~\ref{eq:score} weighs.
We did not refit routings at other loads, but a routing fitted at one load
need not suit another; the routings we report are fitted and measured at 32
concurrent requests.

\subsection{Overhead of Routing and Estimation}
\label{sec:res-overhead}
The router adds one embedding and one nearest-centroid lookup per query, and the
estimator one forward pass of a small encoder per gated answer.
On the reasoning benchmarks, with longer outputs, each adds at most 0.01\% to the
mean E2EL.
The largest share is on TeleQnA, whose outputs are short; there the estimator adds
1.4\% when it scores answers one at a time, and about 0.1\% in batches of 32
(Appendix~\ref{sec:appendix-overhead}).
The overhead is therefore negligible, and the costs we report exclude it, so that
every cost is what the serving stack measured.

\section{Discussion and Limitations}
\label{sec:discussion}

\paragraph{Choosing the cost term.}
A per-token budget is blind to answer length, so it admits models that reason at
length and make every request slow (Section~\ref{sec:res-cost}).
When the models in a pool differ in how long they answer, and in particular
when some reason and others do not, we recommend fitting on E2EL, or on both
and comparing the routings they select.
TPOT remains appropriate where the serving constraint is decoding speed
itself.

\paragraph{Refitting the router.}
Because hardware and load change models' costs unevenly
(Section~\ref{sec:res-hardware}), a deployment that changes either should fit
again.
Refitting is cheap, as the correctness labels do not change, so only the models'
costs need measuring on the new configuration, and an estimator is reused for
any model that stays gated.
Adding a model likewise requires running only that model on the training
queries; no other model is evaluated again and no new annotation is collected.
Pareto pruning then decides whether the new model can ever be selected, and an
estimator is needed only if it ends up gated.

\paragraph{When CRE-Router is not needed.}
Routing pays when no single model is both accurate enough and cheap enough for
the budget; where one model is, an operator may prefer it.
When models are bought from an API priced per token, the cost models of the
prior methods apply as they stand.

\paragraph{Limitations.}
The method needs task-correctness labels, so it targets tasks with verifiable
answers; the estimator consumes any binary correctness signal, but producing
such labels for open-ended tasks, for example with a model as judge, is outside
our scope.
Clusters and routings are fitted offline and do not adapt to a shift in the
query distribution.
The budget bounds the Stage~1 cost measured on training data, so the test set can cost more, as it does on TeleMath under E2EL.
All measurements are on A100 GPUs; other accelerators change models' costs in
ways we have not measured.
The estimator judges whether the efficient model's answer is correct, not
whether the strong model can correct it, so some escalations add the strong
model's cost without correcting the answer.

\paragraph{Future work.}
Among potential directions, Stage~2 could weigh what an escalation costs as well as whether the answer is
wrong, for example by predicting how long the strong model's answer will be, and
could decide earlier, from the opening of the efficient model's reasoning rather
than its finished answer.
Escalation could also learn which model to escalate to, and continue along more
than one tier.
The routing could adapt online as the query distribution shifts, and be
measured on other accelerators and on open-ended tasks labelled by a model as
judge.

\section{Conclusion}
\label{sec:conclusion}

On premises, the cost of a model differs from its cost through an API; it
depends on the hardware that serves it and the load it serves.
CRE-Router fits its routing to that measured cost.
It sends each cluster of queries to the model that best trades error against
cost; a quality estimator then accepts or rejects each answer, and only the
queries whose answers it rejects are escalated to the strong model.
On TeleMath, CRE-Router is comparable in accuracy to the strong model at both
budgets, with a lower TPOT, and significantly more accurate than FrugalGPT under the same
budget.

Our results also hold three lessons for serving routed models.
First, the cost term is a design decision; budgets on TPOT and on E2EL select
different models, because only E2EL sees how long the answers are.
Second, a cluster is a unit of cost as well as of error; pricing each model per
cluster lowers E2EL at comparable accuracy.
Third, a routing belongs to its serving configuration.
One and two A100s select different routings at the same budget, so a change of
hardware calls for refitting, which needs new cost measurements but no new
labels. For an operator serving open-weight models on premises, the cost term, the
budget and the hardware are therefore part of the routing decision, and
CRE-Router fits all three from labels that standard benchmark evaluation
already provides.

\section*{Broader Impact Statement}

This work evaluates publicly available models on publicly available benchmarks.
No personally identifiable information is used.
Routing frameworks that reduce inference cost may lower the barrier to
deploying capable models, with both positive (accessibility and energy savings)
and negative (increased deployment at scale) implications.

\section*{Acknowledgements}
This work was funded by Huawei Ireland through a research collaboration with
the ADAPT Centre at Trinity College Dublin.

\bibliography{paperpile,telecom}
\bibliographystyle{tmlr}

\appendix
\newpage\clearpage
\raggedbottom
\section{Serving Configuration}
\label{sec:appendix-inference}

\paragraph{Software.}
Every reported measurement uses vLLM~0.19.0 \citep{Kwon2023-vLLM}, PyTorch~2.10.0
and Transformers~5.5.3.
The released repository pins these versions.

\paragraph{Server.}
Each model is served alone through vLLM's OpenAI-compatible server.
The server reserves 92\% of GPU memory and uses seed~0.
Its context length is set per model, between 32,768 and 49,152 tokens, so that it
holds the prompt and the output cap.
The key-value cache keeps the model's data type.

\paragraph{Load.}
Requests are sent all at once by vLLM's benchmark client, which keeps at most 32
in progress.

\paragraph{Sampling.}
We use the sampling parameters recommended by the Qwen3 team: temperature 0.6
and top-$p$ 0.95 for reasoning models, and temperature 0.7 and top-$p$ 0.8 for
models answering directly, both with top-$k$ 20 and min-$p$ 0.
For consistency, we apply the same settings to the Gemma~4 and VibeThinker models.
Gemma~4 models are switched between reasoning and direct answers through the
thinking option of their chat template.
Outputs are capped at 40,960 tokens with reasoning, at 16,384 tokens for
TeleMath without reasoning, and at 1,024 tokens for TeleQnA; Qwen3-8B, whose
40,960-token context window cannot hold both a question and the larger cap, is
capped at 38,912 tokens.

\subsection*{Prompts}
\label{sec:appendix-prompts}

Every model receives the question as its user prompt, followed by a
benchmark-specific instruction that fixes the format of the answer.

\paragraph{TeleMath.}
\begin{small}
\begin{verbatim}
\n\nSolve the problem and give the final numerical answer, in the unit stated in the question,
inside \boxed{}.
\end{verbatim}
\end{small}

\paragraph{AIME.}
\begin{small}
\begin{verbatim}
\n\nThink briefly about the approach, then provide your final answer as a number.
\end{verbatim}
\end{small}

\paragraph{TeleQnA.}
The instruction precedes the options, which are those of each question; the
example below shows one.
\begin{small}
\begin{verbatim}
\nProvide the answer in the following format:
Explanation: <one_sentence_explanation>
Answer: <choice_number_only>
0. F1 interface
1. E2 interface
2. CPRI interface
3. O1 interface
4. A1 interface
\end{verbatim}
\end{small}

\section{Estimator Training}
\label{sec:appendix-qe-classifiers}

Each quality estimator is ModernBERT-base \citep{Warner2025-ModernBERT}
fine-tuned as a binary classifier on the answers of the model it gates.
An answer is labelled \textit{accept} when it is correct and \textit{escalate}
otherwise, the same task-correctness labels Stage~1 uses.
The input is the query, the answer and the answer's length in tokens, joined by
the separator token.
The reasoning trace is removed first, so the estimator judges the answer the
model delivers.
Answers are then truncated to their last 1,000 words, which keeps the final
answer, and the input to 4,096 tokens, or 512 for TeleQnA, whose answers are
short.
The length is that of the whole generation, reasoning included.
Table~\ref{tab:appendix-estimators} lists every estimator a reported result
uses, with its training data and settings.

\begin{table}[ht]
\centering
\caption{\textbf{Quality estimators.} The quality estimators behind every reported Stage~2 result. Each is ModernBERT-base fine-tuned on one GPU on the gated model's own answers to the training questions from five runs, labelled by task correctness, with class-weighted cross-entropy, batch size 64 and up to ten epochs. Training stops early, with patience 4, and keeps the checkpoint with the best macro-F1 on the model's answers to the test questions from the first run (all five runs on AIME). Served on is the hardware that generated the answers an estimator scores, not the hardware the estimator runs on.}
\label{tab:appendix-estimators}
\small
\begin{adjustbox}{max width=\textwidth}
\begin{tabular}{llllrrrr}
\toprule
\textbf{Benchmark} & \textbf{Gated model} & \textbf{Gated clusters} & \textbf{Served on} & \textbf{Train answers} & \textbf{Early-stopping answers} & \textbf{LR} & \textbf{Max length} \\
\midrule
TeleMath & Gemma4-E2B & C1 & one A100 & 1,495 & 201 & 2e-5 & 4,096 \\
 & Gemma4-E4B-think & C3 & one A100 & 1,495 & 201 & 2e-5 & 4,096 \\
 & Gemma4-26B & C2 and C3 & one A100 & 1,495 & 201 & 2e-5 & 4,096 \\
 & Qwen3-30B-think & C0 and C1 & two A100s & 1,495 & 201 & 2e-5 & 4,096 \\
 & Gemma4-E2B & $k = 1$ ablation, E2EL & one A100 & 1,495 & 201 & 2e-5 & 4,096 \\
 & Qwen3-4B-Instruct & $k = 1$ ablation, TPOT & one A100 & 1,495 & 201 & 2e-5 & 4,096 \\
\midrule
AIME & VibeThinker-1.5B & C1 & one and two A100s & 4,605 & 450 & 5e-5 & 4,096 \\
\midrule
TeleQnA & Qwen3-4B-Instruct & C0 & one and two A100s & 45,000 & 1,000 & 2e-5 & 512 \\
 & Gemma4-E2B & C0, two-model comparison & one A100 & 45,000 & 1,000 & 2e-5 & 512 \\
\bottomrule
\end{tabular}
\end{adjustbox}
\end{table}

\section{Routings at Every Budget}
\label{sec:appendix-routings}

Table~\ref{tab:appendix-ladder} lists every routing CRE-Router can select on
TeleMath under each cost term, from the most accurate to the cheapest, and marks
the ones the reported budgets select.
An operator with a different budget reads its routing from this table without
fitting again.

\begin{table}[t]
\centering
\caption{\textbf{\boldmath$\lambda$ sweep on TeleMath.} Every routing CRE-Router can select on TeleMath, one A100, under each cost term: the regions of $\lambda$ over which the cluster assignment of the Pareto-pruned pool stays the same, in order of increasing $\lambda$. Accuracy and cost are Stage~1 on the training set, the values the budget is fitted against; a budget selects the most accurate routing whose training cost is within it. The last column marks the routings the reported budgets select, the only ones measured on the test set (Tables~\ref{tab:telemath-main} and~\ref{tab:telemath-e2el}).}
\label{tab:appendix-ladder}
\small
\begin{adjustbox}{max width=\textwidth}
\begin{tabular}{llllllrrl}
\toprule
\textbf{Cost term} & \textbf{$\lambda$} & \textbf{C0} & \textbf{C1} & \textbf{C2} & \textbf{C3} & \textbf{Acc.\ (\%)} & \textbf{Cost} & \textbf{Selected by} \\
\midrule
TPOT, cost in ms & $[0.000, 0.070)$ & Qwen3-30B-think & Qwen3-30B-think & Gemma4-26B-think & Gemma4-26B-think & 74.2 & 27.1 &  \\
 & $[0.070, 0.216)$ & Qwen3-30B-think & Qwen3-30B-think & Qwen3-30B-think & Gemma4-26B-think & 73.8 & 26.2 &  \\
 & $[0.216, 0.313)$ & Qwen3-30B-think & Gemma4-E2B & Qwen3-30B-think & Gemma4-26B-think & 69.1 & 20.9 & $B = 25$\,ms \\
 & $[0.313, 0.467)$ & Qwen3-30B-think & Gemma4-E2B & Qwen3-30B-think & Gemma4-E4B-think & 64.7 & 17.5 & $B = 20$\,ms \\
 & $[0.467, 0.515)$ & Qwen3-4B-Instruct & Gemma4-E2B & Qwen3-30B-think & Gemma4-E4B-think & 57.5 & 13.6 &  \\
 & $[0.515, 0.689)$ & Qwen3-4B-Instruct & Gemma4-E2B & Qwen3-4B-Instruct & Gemma4-E4B-think & 53.3 & 11.3 &  \\
 & $[0.689, 0.704)$ & Gemma4-E2B & Gemma4-E2B & Qwen3-4B-Instruct & Gemma4-E4B-think & 49.0 & 9.2 &  \\
 & $[0.704, 0.923)$ & Gemma4-E2B & Gemma4-E2B & Qwen3-4B-Instruct & Gemma4-E2B-think & 47.0 & 8.5 &  \\
 & $[0.923, 1.609)$ & Gemma4-E2B & Gemma4-E2B & Gemma4-E2B-think & Gemma4-E2B-think & 44.3 & 7.7 &  \\
 & $[1.609, 3.945)$ & Gemma4-E2B & Gemma4-E2B & Gemma4-E2B & Gemma4-E2B-think & 43.7 & 7.6 &  \\
 & $[3.945, \infty)$ & Gemma4-E2B & Gemma4-E2B & Gemma4-E2B & Gemma4-E2B & 42.4 & 7.6 &  \\
\midrule
E2EL, cost in s & $[0.000, 0.025)$ & Qwen3-30B-think & Qwen3-30B-think & Gemma4-26B-think & Gemma4-26B-think & 74.2 & 255.1 &  \\
 & $[0.025, 0.120)$ & Qwen3-30B-think & Qwen3-30B-think & Gemma4-26B & Gemma4-26B-think & 73.8 & 178.4 &  \\
 & $[0.120, 0.182)$ & Qwen3-30B-think & Qwen3-30B-think & Gemma4-26B & Gemma4-26B & 71.8 & 113.7 & $B = 120$\,s \\
 & $[0.182, 0.188)$ & Qwen3-30B-think & Gemma4-26B & Gemma4-26B & Gemma4-26B & 69.8 & 84.6 &  \\
 & $[0.188, 1.972)$ & Gemma4-26B & Gemma4-26B & Gemma4-26B & Gemma4-26B & 67.7 & 29.4 &  \\
 & $[1.972, 6.157)$ & Gemma4-26B & Gemma4-E2B & Gemma4-26B & Gemma4-26B & 65.0 & 23.5 & $B = 25$\,s \\
 & $[6.157, 7.648)$ & Gemma4-E4B & Gemma4-E2B & Gemma4-26B & Gemma4-26B & 59.1 & 19.2 &  \\
 & $[7.648, 8.849)$ & Gemma4-E4B & Gemma4-E2B & Gemma4-26B & Gemma4-E4B & 55.3 & 17.8 &  \\
 & $[8.849, 9.125)$ & Gemma4-E2B & Gemma4-E2B & Gemma4-26B & Gemma4-E4B & 51.7 & 15.7 &  \\
 & $[9.125, 10.422)$ & Gemma4-E2B & Gemma4-E2B & Gemma4-26B & Gemma4-E2B & 49.8 & 15.1 &  \\
 & $[10.422, \infty)$ & Gemma4-E2B & Gemma4-E2B & Gemma4-E2B & Gemma4-E2B & 42.4 & 12.0 &  \\
\bottomrule
\end{tabular}
\end{adjustbox}
\end{table}

\section{Baselines in Detail}
\label{sec:appendix-baselines}

\subsection{FrugalGPT Across Budgets}
\label{sec:appendix-frugal-sweep}

\begin{figure}[t]
\centering
\includegraphics[width=\linewidth]{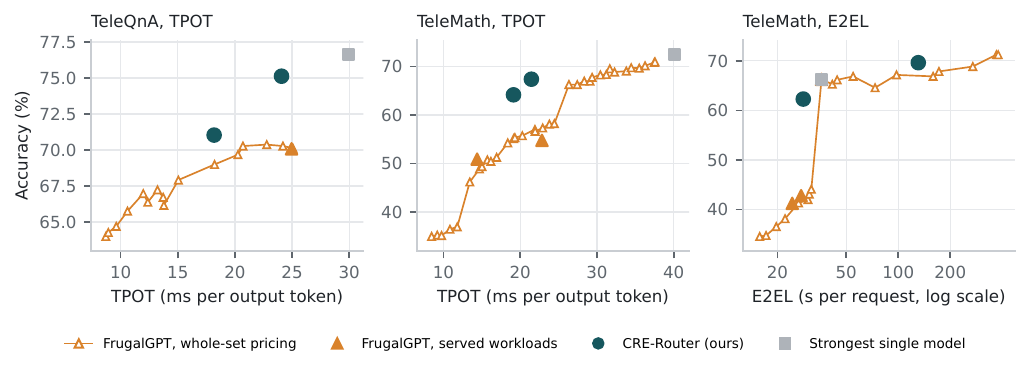}
\caption{\textbf{FrugalGPT across budgets.} FrugalGPT fitted on the training split at every budget on a grid and applied to the test split unchanged, on TeleQnA (TPOT) and TeleMath (TPOT and E2EL), one A100. Hollow triangles are priced on each model's whole-test-set run, which charges an escalated question the cost it has when the whole test set is served; filled triangles are the budgets reported in the main text, priced on the workloads FrugalGPT served. At those budgets the two pricings differ by up to 3.5\,ms per token and 3.2\,s per request. CRE-Router (circles) and the strongest single model (square) are priced on the workloads they served. Higher and further left is better.}
\label{fig:frugal-sweep}
\end{figure}

In Section~\ref{sec:res-baselines} FrugalGPT is given the cost CRE-Router spends at
each reported budget.
Figure~\ref{fig:frugal-sweep} fits it at every budget on a grid instead, with the
same scorers, lists and threshold search.

\subsection{HybridLLM Label Schemes and Pair}
\label{sec:appendix-hybrid}

Table~\ref{tab:appendix-hybrid} gives HybridLLM under its deterministic and
probabilistic label schemes at every budget we report.
The choice of pair matters little.
With Gemma4-E4B-think as the small model in place of Gemma4-E2B, HybridLLM's
probabilistic router is 1.4 points more accurate at $B = 20$\,ms and 0.7 points
less accurate at 25\,ms, with both pairs scored on the same answers.

\begin{table}[t]
\centering
\caption{\textbf{HybridLLM label schemes.} HybridLLM on TeleMath, one A100, under its deterministic and probabilistic label schemes \citep{Ding2024-HybridLLM}, at every budget we report; with 0/1 correctness the transformed labels equal the probabilistic ones, so the two schemes shown are the only distinct ones. Each scheme's router is fitted on the training set, its threshold chosen as the most accurate within the budget, and the two subsets it sends to the small and the large model are served as their own workloads and measured as in Table~\ref{tab:telemath-main}. The probabilistic router, marked with $\ast$, is the one the main tables show. At every budget it is the more accurate of the two.}
\label{tab:appendix-hybrid}
\small
\begin{adjustbox}{max width=\textwidth}
\begin{tabular}{lllrrrr}
\toprule
\textbf{Budget} & \textbf{Models} & \textbf{Scheme} & \textbf{To small (\%)} & \textbf{Acc.\ (\%)} & \textbf{TPOT (ms)} & \textbf{E2EL (s)} \\
\midrule
TPOT 20\,ms & Gemma4-E2B, Qwen3-30B-think & deterministic & 49.3 & 55.5 & 17.7 & 118.4 \\
 &  & probabilistic$^\ast$ & 45.8 & 56.8 & 19.3 & 149.2 \\
\midrule
TPOT 25\,ms & Gemma4-E2B, Qwen3-30B-think & deterministic & 26.4 & 62.1 & 23.2 & 171.2 \\
 &  & probabilistic$^\ast$ & 23.9 & 63.4 & 24.7 & 193.1 \\
\midrule
E2EL 25\,s & Gemma4-E2B, Gemma4-26B & deterministic & 40.8 & 56.4 & 18.5 & 29.0 \\
 &  & probabilistic$^\ast$ & 34.3 & 60.4 & 19.7 & 29.3 \\
\bottomrule
\end{tabular}
\end{adjustbox}
\end{table}

\section{Paired Comparisons}
\label{sec:appendix-comparisons}

Tables~\ref{tab:comparisons} and~\ref{tab:comparisons-pool} list every comparison
the main text describes as significant or comparable, with the interval behind it.

\begin{table}[p]
\centering
\caption{\textbf{Paired comparisons.} The 95\% interval behind every comparison the text makes, so each claim of a significant difference, or of none, can be checked here. Each row is CRE-Router, or the variant named, minus the system it is compared with, on the same questions (Section~\ref{sec:statistics}); a difference is significant when its interval excludes zero. Accuracy in percentage points, TPOT in milliseconds, E2EL in seconds; a negative cost difference means CRE-Router is cheaper. Group headings name the section that makes the claims. The prior methods were run on TeleMath and TeleQnA, the ablations, the cost-term comparison and per-cluster cost on TeleMath only, and the hardware comparison covers all three benchmarks. The estimator's comparisons are in Table~\ref{tab:stage2}.}
\label{tab:comparisons}
\footnotesize
\begin{adjustbox}{max width=\textwidth}
\begin{tabular}{lllllll}
\toprule
\textbf{Benchmark} & \textbf{GPUs} & \textbf{Against} & \textbf{Budget} & \textbf{$\Delta$ Acc.\ (pp)} & \textbf{$\Delta$ TPOT (ms)} & \textbf{$\Delta$ E2EL (s)} \\
\midrule
\multicolumn{7}{l}{\textit{Single models} (Section~\ref{sec:res-single})} \\
TeleMath & one A100 & Qwen3-30B-think & TPOT 20\,ms & $-3.1$ {\scriptsize [$-8.3$, $+2.0$]} & $-9.9$ {\scriptsize [$-11.0$, $-8.7$]} & $-74.6$ {\scriptsize [$-88.5$, $-61.1$]} \\
TeleMath & one A100 & Qwen3-30B-think & TPOT 25\,ms & $+0.1$ {\scriptsize [$-5.1$, $+5.4$]} & $-7.5$ {\scriptsize [$-8.8$, $-6.3$]} & $-19.6$ {\scriptsize [$-39.8$, $+1.5$]} \\
TeleMath & one A100 & Qwen3-8B & TPOT 20\,ms & $+9.5$ {\scriptsize [$+3.5$, $+15.6$]} & $-21.9$ {\scriptsize [$-23.9$, $-20.0$]} & $-377.8$ {\scriptsize [$-415.5$, $-342.8$]} \\
TeleMath & one A100 & Qwen3-8B & TPOT 25\,ms & $+12.8$ {\scriptsize [$+6.6$, $+18.9$]} & $-19.6$ {\scriptsize [$-21.6$, $-17.6$]} & $-322.7$ {\scriptsize [$-364.6$, $-282.3$]} \\
TeleMath & one A100 & Gemma4-26B-think & TPOT 20\,ms & $-8.3$ {\scriptsize [$-14.1$, $-2.5$]} & $-21.0$ {\scriptsize [$-22.2$, $-19.7$]} & $-424.9$ {\scriptsize [$-461.5$, $-392.3$]} \\
TeleMath & one A100 & Gemma4-26B-think & TPOT 25\,ms & $-5.1$ {\scriptsize [$-10.3$, $+0.1$]} & $-18.7$ {\scriptsize [$-20.0$, $-17.3$]} & $-369.9$ {\scriptsize [$-408.0$, $-334.7$]} \\
TeleMath & one A100 & Gemma4-26B & TPOT 20\,ms & $-2.1$ {\scriptsize [$-7.5$, $+3.1$]} & $-7.2$ {\scriptsize [$-8.3$, $-6.2$]} & $+103.0$ {\scriptsize [$+87.5$, $+118.7$]} \\
TeleMath & one A100 & Gemma4-26B & TPOT 25\,ms & $+1.1$ {\scriptsize [$-4.0$, $+6.1$]} & $-4.9$ {\scriptsize [$-6.0$, $-3.8$]} & $+158.0$ {\scriptsize [$+137.4$, $+179.3$]} \\
\midrule
\multicolumn{7}{l}{\textit{Prior methods, TPOT budget} (Section~\ref{sec:res-baselines})} \\
TeleMath & one A100 & FrugalGPT & TPOT 20\,ms & $+13.3$ {\scriptsize [$+7.3$, $+19.4$]} & $+4.7$ {\scriptsize [$+3.7$, $+5.7$]} & $+75.4$ {\scriptsize [$+58.7$, $+92.2$]} \\
TeleMath & one A100 & FrugalGPT & TPOT 25\,ms & $+12.6$ {\scriptsize [$+6.9$, $+18.4$]} & $-1.4$ {\scriptsize [$-3.9$, $+1.1$]} & $+149.8$ {\scriptsize [$+129.7$, $+170.6$]} \\
TeleMath & one A100 & HybridLLM, two models, Stage~1 & TPOT 20\,ms & $+5.5$ {\scriptsize [$+0.2$, $+10.8$]} & $-0.5$ {\scriptsize [$-1.8$, $+0.9$]} & $-13.0$ {\scriptsize [$-30.5$, $+3.8$]} \\
TeleMath & one A100 & HybridLLM, two models, Stage~1 & TPOT 25\,ms & $+3.9$ {\scriptsize [$-0.9$, $+8.7$]} & $+4.3$ {\scriptsize [$+2.7$, $+5.9$]} & $+20.6$ {\scriptsize [$+2.9$, $+37.5$]} \\
TeleMath & one A100 & HybridLLM, two models, Stage~1+2 & TPOT 20\,ms & $+8.2$ {\scriptsize [$+3.1$, $+13.4$]} & $+0.6$ {\scriptsize [$-0.8$, $+1.9$]} & $-1.2$ {\scriptsize [$-18.3$, $+15.3$]} \\
TeleMath & one A100 & HybridLLM, full pool, Stage~1 & TPOT 25\,ms & $+1.3$ {\scriptsize [$-3.8$, $+6.4$]} & $-4.3$ {\scriptsize [$-5.7$, $-2.8$]} & $-10.8$ {\scriptsize [$-33.9$, $+12.3$]} \\
TeleQnA & one A100 & FrugalGPT & TPOT 20\,ms & $+5.0$ {\scriptsize [$+3.0$, $+7.1$]} & $-0.9$ {\scriptsize [$-2.1$, $+0.3$]} & $+0.0$ {\scriptsize [$-0.0$, $+0.1$]} \\
TeleQnA & one A100 & HybridLLM, two models, Stage~1 & TPOT 20\,ms & $-2.3$ {\scriptsize [$-4.3$, $-0.4$]} & $-2.1$ {\scriptsize [$-2.9$, $-1.3$]} & $-0.1$ {\scriptsize [$-0.2$, $-0.1$]} \\
TeleQnA & one A100 & HybridLLM, two models, Stage~1+2 & TPOT 20\,ms & $+2.7$ {\scriptsize [$+0.8$, $+4.5$]} & $+4.5$ {\scriptsize [$+3.7$, $+5.3$]} & $+0.2$ {\scriptsize [$+0.2$, $+0.3$]} \\
TeleQnA & one A100 & HybridLLM, full pool, Stage~1 & TPOT 20\,ms & $-1.2$ {\scriptsize [$-3.6$, $+1.2$]} & $-0.9$ {\scriptsize [$-1.7$, $-0.1$]} & $-0.1$ {\scriptsize [$-0.1$, $-0.0$]} \\
\midrule
\multicolumn{7}{l}{\textit{Prior methods, E2EL budget} (Section~\ref{sec:res-baselines})} \\
TeleMath & one A100 & FrugalGPT & E2EL 25\,s & $+19.5$ {\scriptsize [$+13.5$, $+25.7$]} & $+7.5$ {\scriptsize [$+5.7$, $+9.4$]} & $+0.8$ {\scriptsize [$-3.4$, $+4.7$]} \\
TeleMath & one A100 & HybridLLM (Stage~1) & E2EL 25\,s & $-0.4$ {\scriptsize [$-4.6$, $+3.8$]} & $-0.3$ {\scriptsize [$-1.4$, $+0.7$]} & $-3.4$ {\scriptsize [$-7.4$, $-0.3$]} \\
TeleMath & one A100 & No-clustering ablation & E2EL 25\,s & $+0.7$ {\scriptsize [$-3.3$, $+4.8$]} & $-5.9$ {\scriptsize [$-8.2$, $-3.7$]} & $-13.5$ {\scriptsize [$-17.4$, $-9.8$]} \\
TeleMath & one A100 & Gemma4-26B & E2EL 25\,s & $-4.0$ {\scriptsize [$-7.8$, $-0.3$]} & $-4.7$ {\scriptsize [$-5.8$, $-3.4$]} & $-7.8$ {\scriptsize [$-12.5$, $-3.9$]} \\
\midrule
\multicolumn{7}{l}{\textit{Ablations} (Section~\ref{sec:res-k1})} \\
TeleMath & one A100 & No-clustering ablation & TPOT 20\,ms & $+4.0$ {\scriptsize [$-1.1$, $+9.1$]} & $-9.9$ {\scriptsize [$-11.9$, $-8.1$]} & $-70.1$ {\scriptsize [$-94.1$, $-48.9$]} \\
TeleMath & one A100 & No-clustering ablation & TPOT 25\,ms & $+0.1$ {\scriptsize [$-5.1$, $+5.4$]} & $-7.5$ {\scriptsize [$-8.8$, $-6.3$]} & $-19.6$ {\scriptsize [$-39.9$, $+1.4$]} \\
TeleMath & one A100 & Stage~1+2 $-$ Stage~1 & TPOT 20\,ms & $+5.3$ {\scriptsize [$+2.1$, $+8.6$]} & $+1.8$ {\scriptsize [$+1.2$, $+2.4$]} & $+21.3$ {\scriptsize [$+14.2$, $+29.3$]} \\
TeleMath & one A100 & Stage~1+2 $-$ Stage~1 & TPOT 25\,ms & $+2.7$ {\scriptsize [$+0.3$, $+5.5$]} & $+1.0$ {\scriptsize [$+0.6$, $+1.5$]} & $+11.8$ {\scriptsize [$+6.3$, $+18.0$]} \\
TeleMath & one A100 & full pool $-$ two models, Stage~1 & TPOT 20\,ms & $-3.4$ {\scriptsize [$-7.8$, $+1.0$]} & $-1.5$ {\scriptsize [$-2.0$, $-1.0$]} & $-18.5$ {\scriptsize [$-28.8$, $-8.1$]} \\
TeleMath & one A100 & full pool $-$ two models, Stage~1 & TPOT 25\,ms & $-2.6$ {\scriptsize [$-8.1$, $+3.0$]} & $-8.6$ {\scriptsize [$-9.9$, $-7.2$]} & $-31.4$ {\scriptsize [$-53.5$, $-8.6$]} \\
TeleMath & one A100 & full pool $-$ two models, Stage~1+2 & TPOT 20\,ms & $-0.8$ {\scriptsize [$-4.7$, $+3.0$]} & $-0.7$ {\scriptsize [$-1.2$, $-0.2$]} & $-8.9$ {\scriptsize [$-18.1$, $+0.7$]} \\
TeleMath & one A100 & full pool $-$ two models, Stage~1+2 & TPOT 25\,ms & $+0.1$ {\scriptsize [$-5.1$, $+5.4$]} & $-7.5$ {\scriptsize [$-8.8$, $-6.3$]} & $-19.6$ {\scriptsize [$-39.8$, $+1.5$]} \\
\midrule
\multicolumn{7}{l}{\textit{Looser E2EL budget} (Section~\ref{sec:res-cost})} \\
TeleMath & one A100 & Qwen3-30B-think & E2EL 120\,s & $+2.4$ {\scriptsize [$-2.1$, $+7.1$]} & $-5.6$ {\scriptsize [$-6.2$, $-5.0$]} & $-83.1$ {\scriptsize [$-99.3$, $-67.4$]} \\
TeleMath & one A100 & CRE-Router, TPOT 25\,ms & E2EL 120\,s & $+2.2$ {\scriptsize [$-2.5$, $+7.0$]} & $+1.9$ {\scriptsize [$+0.9$, $+3.0$]} & $-63.5$ {\scriptsize [$-87.2$, $-41.1$]} \\
TeleMath & one A100 & CRE-Router, TPOT 20\,ms & E2EL 120\,s & $+5.5$ {\scriptsize [$+0.8$, $+10.2$]} & $+4.3$ {\scriptsize [$+3.3$, $+5.2$]} & $-8.5$ {\scriptsize [$-22.4$, $+5.3$]} \\
TeleMath & one A100 & CRE-Router, E2EL 25\,s & E2EL 120\,s & $+7.3$ {\scriptsize [$+2.6$, $+12.1$]} & $+1.7$ {\scriptsize [$+0.5$, $+2.9$]} & $+102.4$ {\scriptsize [$+88.9$, $+116.5$]} \\
\midrule
\multicolumn{7}{l}{\textit{Per-cluster cost} (Section~\ref{sec:res-percluster})} \\
TeleMath & one A100 & per-cluster $-$ per-model & TPOT 20\,ms & $+0.6$ {\scriptsize [$-3.4$, $+4.7$]} & $+0.7$ {\scriptsize [$+0.1$, $+1.2$]} & $-30.3$ {\scriptsize [$-46.0$, $-15.3$]} \\
TeleMath & one A100 & per-cluster $-$ per-model & TPOT 25\,ms & $-0.2$ {\scriptsize [$-3.6$, $+3.3$]} & $-0.0$ {\scriptsize [$-0.4$, $+0.3$]} & $-39.2$ {\scriptsize [$-53.5$, $-25.5$]} \\
\midrule
\multicolumn{7}{l}{\textit{Hardware, Stage~1+2 against single models} (Section~\ref{sec:res-hardware})} \\
AIME & two A100s & Qwen3-Next-80B-A3B-Thinking & TPOT 20\,ms & $-1.0$ {\scriptsize [$-5.6$, $+3.2$]} & $-5.5$ {\scriptsize [$-6.6$, $-4.3$]} & $-111.4$ {\scriptsize [$-137.9$, $-85.6$]} \\
TeleMath & two A100s & Qwen3-Next-80B-A3B-Thinking & TPOT 20\,ms & $+3.6$ {\scriptsize [$-0.8$, $+8.1$]} & $+0.8$ {\scriptsize [$+0.1$, $+1.5$]} & $+90.2$ {\scriptsize [$+73.1$, $+108.1$]} \\
TeleQnA & two A100s & Qwen3-Next-80B-A3B-Instruct & TPOT 15\,ms & $+2.0$ {\scriptsize [$-0.5$, $+4.4$]} & $-19.5$ {\scriptsize [$-20.0$, $-19.0$]} & $-1.1$ {\scriptsize [$-1.2$, $-1.1$]} \\
TeleQnA & two A100s & Qwen3-Next-80B-A3B-Thinking & TPOT 15\,ms & $-2.0$ {\scriptsize [$-4.4$, $+0.3$]} & $-0.7$ {\scriptsize [$-1.1$, $-0.3$]} & $-23.7$ {\scriptsize [$-25.2$, $-22.3$]} \\
TeleMath & one A100 & Qwen3-8B & TPOT 25\,ms & $+12.8$ {\scriptsize [$+6.6$, $+18.9$]} & $-19.6$ {\scriptsize [$-21.6$, $-17.6$]} & $-322.7$ {\scriptsize [$-364.6$, $-282.3$]} \\
TeleMath & two A100s & Qwen3-8B & TPOT 20\,ms & $+18.9$ {\scriptsize [$+13.3$, $+24.6$]} & $-2.3$ {\scriptsize [$-3.3$, $-1.3$]} & $-38.3$ {\scriptsize [$-58.5$, $-18.1$]} \\
TeleQnA & one A100 & Qwen3-8B & TPOT 20\,ms & $+1.7$ {\scriptsize [$-0.6$, $+4.1$]} & $+8.2$ {\scriptsize [$+7.5$, $+8.9$]} & $-12.9$ {\scriptsize [$-13.6$, $-12.3$]} \\
TeleQnA & two A100s & Qwen3-8B & TPOT 15\,ms & $+1.4$ {\scriptsize [$-1.0$, $+3.7$]} & $+7.2$ {\scriptsize [$+6.8$, $+7.6$]} & $-7.8$ {\scriptsize [$-8.2$, $-7.4$]} \\
AIME & one A100 & Qwen3-8B & TPOT 30\,ms & $+14.6$ {\scriptsize [$+7.8$, $+22.0$]} & $-21.9$ {\scriptsize [$-25.4$, $-18.7$]} & $-348.3$ {\scriptsize [$-436.8$, $-264.8$]} \\
AIME & two A100s & Qwen3-8B & TPOT 20\,ms & $+17.2$ {\scriptsize [$+10.8$, $+24.2$]} & $-9.6$ {\scriptsize [$-11.2$, $-8.0$]} & $-158.7$ {\scriptsize [$-198.6$, $-120.6$]} \\
\bottomrule
\end{tabular}
\end{adjustbox}
\end{table}
\begin{table}[t]
\centering
\caption{\textbf{Paired comparisons on AIME and TeleQnA.} CRE-Router (Stage~1+2) minus each pool model serving every question alone, on one and on two A100s, the comparisons Sections~\ref{sec:res-single} and~\ref{sec:res-baselines} make for these benchmarks, read as Table~\ref{tab:comparisons}.}
\label{tab:comparisons-pool}
\footnotesize
\begin{adjustbox}{max width=\textwidth}
\begin{tabular}{lllllll}
\toprule
\textbf{Benchmark} & \textbf{GPUs} & \textbf{Against} & \textbf{Budget} & \textbf{$\Delta$ Acc.\ (pp)} & \textbf{$\Delta$ TPOT (ms)} & \textbf{$\Delta$ E2EL (s)} \\
\midrule
AIME & one A100 & VibeThinker-1.5B & TPOT 30\,ms & $+11.3$ {\scriptsize [$+4.8$, $+18.3$]} & $+9.3$ {\scriptsize [$+7.4$, $+11.3$]} & $+174.8$ {\scriptsize [$+129.6$, $+222.6$]} \\
AIME & one A100 & Qwen3-30B-think & TPOT 30\,ms & $+0.2$ {\scriptsize [$-4.9$, $+5.4$]} & $-16.9$ {\scriptsize [$-19.3$, $-14.6$]} & $-304.2$ {\scriptsize [$-368.5$, $-245.1$]} \\
AIME & two A100s & VibeThinker-1.5B & TPOT 20\,ms & $+14.1$ {\scriptsize [$+7.8$, $+21.0$]} & $+6.8$ {\scriptsize [$+5.7$, $+7.9$]} & $+131.9$ {\scriptsize [$+104.0$, $+161.1$]} \\
AIME & two A100s & Qwen3-30B-think & TPOT 20\,ms & $+2.5$ {\scriptsize [$-2.0$, $+7.1$]} & $-9.4$ {\scriptsize [$-10.8$, $-8.1$]} & $-160.5$ {\scriptsize [$-194.9$, $-128.0$]} \\
\midrule
TeleQnA & one A100 & Qwen3-4B-Instruct & TPOT 20\,ms & $+8.3$ {\scriptsize [$+6.0$, $+10.7$]} & $+13.4$ {\scriptsize [$+12.8$, $+14.1$]} & $+0.7$ {\scriptsize [$+0.7$, $+0.7$]} \\
TeleQnA & one A100 & Gemma4-E2B & TPOT 20\,ms & $+11.5$ {\scriptsize [$+8.7$, $+14.4$]} & $+15.7$ {\scriptsize [$+15.0$, $+16.3$]} & $+0.8$ {\scriptsize [$+0.8$, $+0.9$]} \\
TeleQnA & one A100 & Gemma4-E4B & TPOT 20\,ms & $+7.3$ {\scriptsize [$+4.8$, $+9.8$]} & $+10.4$ {\scriptsize [$+9.7$, $+11.0$]} & $+0.6$ {\scriptsize [$+0.6$, $+0.7$]} \\
TeleQnA & one A100 & Gemma4-26B & TPOT 20\,ms & $-1.5$ {\scriptsize [$-3.0$, $+0.0$]} & $-5.9$ {\scriptsize [$-6.6$, $-5.2$]} & $-0.3$ {\scriptsize [$-0.3$, $-0.3$]} \\
TeleQnA & two A100s & Qwen3-4B-Instruct & TPOT 15\,ms & $+7.5$ {\scriptsize [$+5.1$, $+9.9$]} & $+8.2$ {\scriptsize [$+7.7$, $+8.6$]} & $+0.4$ {\scriptsize [$+0.4$, $+0.5$]} \\
TeleQnA & two A100s & Gemma4-E2B & TPOT 15\,ms & $+10.8$ {\scriptsize [$+7.9$, $+13.7$]} & $+6.0$ {\scriptsize [$+5.5$, $+6.4$]} & $+0.4$ {\scriptsize [$+0.4$, $+0.4$]} \\
TeleQnA & two A100s & Gemma4-E4B & TPOT 15\,ms & $+7.3$ {\scriptsize [$+4.8$, $+9.9$]} & $+4.6$ {\scriptsize [$+4.2$, $+5.1$]} & $+0.3$ {\scriptsize [$+0.3$, $+0.4$]} \\
TeleQnA & two A100s & Gemma4-26B & TPOT 15\,ms & $-1.9$ {\scriptsize [$-3.5$, $-0.4$]} & $-2.7$ {\scriptsize [$-3.1$, $-2.3$]} & $-0.1$ {\scriptsize [$-0.1$, $-0.1$]} \\
\bottomrule
\end{tabular}
\end{adjustbox}
\end{table}

\section{Pricing Basis}
\label{sec:appendix-basis}
\begin{table}[t]
\centering
\caption{\textbf{Cost on two pricing bases.} The TeleMath systems of Tables~\ref{tab:telemath-main} and~\ref{tab:telemath-e2el}, one A100, Stage~1+2, with Stage~1 priced per cluster, each test cluster served as its own workload, and on the whole test set served as one workload. Only costs change; every answer and escalation decision is kept, so accuracy is the same on both bases, and escalation sets are priced as served. A dagger marks the basis the main tables use. Served per cluster, each single model costs 5 to 35\% less per token.}
\label{tab:appendix-basis}
\small
\begin{adjustbox}{max width=\textwidth}
\begin{tabular}{llrrrrr}
\toprule
 & & & \multicolumn{2}{c}{\textbf{TPOT (ms)}} & \multicolumn{2}{c}{\textbf{E2EL (s)}} \\
\cmidrule(lr){4-5}\cmidrule(lr){6-7}
\textbf{System} & \textbf{Budget} & \textbf{Acc.\ (\%)} & \textbf{Per cluster} & \textbf{Whole set} & \textbf{Per cluster} & \textbf{Whole set} \\
\midrule
CRE-Router & TPOT, $B = 20$\,ms & 64.2 & 19.1$^\dagger$ & 22.0 & 139.1$^\dagger$ & 165.3 \\
 & TPOT, $B = 25$\,ms & 67.4 & 21.5$^\dagger$ & 25.6 & 194.1$^\dagger$ & 239.6 \\
 & E2EL, $B = 25$\,s & 62.3 & 21.7$^\dagger$ & 23.6 & 28.2$^\dagger$ & 33.7 \\
 & E2EL, $B = 120$\,s & 69.6 & 23.4$^\dagger$ & 27.4 & 130.6$^\dagger$ & 159.6 \\
\midrule
No-clustering ablation ($k = 1$) & TPOT, $B = 20$\,ms & 60.2 & 24.5 & 29.1$^\dagger$ & 188.9 & 209.2$^\dagger$ \\
 & TPOT, $B = 25$\,ms & 67.3 & 23.1 & 29.0$^\dagger$ & 167.5 & 213.7$^\dagger$ \\
 & E2EL, $B = 25$\,s & 61.6 & 26.4 & 27.6$^\dagger$ & 39.6 & 41.7$^\dagger$ \\
\midrule
Qwen3-30B-think & -- & 67.3 & 23.1 & 29.0$^\dagger$ & 167.5 & 213.7$^\dagger$ \\
Qwen3-4B-think & -- & 47.6 & 22.1 & 32.5$^\dagger$ & 245.7 & 351.2$^\dagger$ \\
Gemma4-26B-think & -- & 72.4 & 31.1 & 40.1$^\dagger$ & 429.5 & 564.0$^\dagger$ \\
Gemma4-26B & -- & 66.3 & 24.0 & 26.3$^\dagger$ & 29.3 & 36.0$^\dagger$ \\
Gemma4-E4B-think & -- & 45.8 & 12.2 & 13.1$^\dagger$ & 32.9 & 35.1$^\dagger$ \\
Gemma4-E4B & -- & 45.3 & 11.6 & 12.3$^\dagger$ & 17.6 & 18.4$^\dagger$ \\
Qwen3-4B-Instruct & -- & 42.2 & 11.7 & 18.1$^\dagger$ & 38.8 & 59.2$^\dagger$ \\
Gemma4-E2B-think & -- & 37.8 & 7.9 & 8.7$^\dagger$ & 23.4 & 26.3$^\dagger$ \\
Gemma4-E2B & -- & 34.0 & 7.4 & 8.1$^\dagger$ & 12.7 & 14.4$^\dagger$ \\
\bottomrule
\end{tabular}
\end{adjustbox}
\end{table}

Each system is priced on the workloads it serves (Section~\ref{sec:measured-cost}).
A routed Stage~1 is priced on each test cluster served as its own workload, and a
single model, or the Stage~1 of the no-clustering ablation, on the whole test set
served as one workload.
Served per cluster, the same model answering the same questions costs between 5
and 35\% less per token on TeleMath (Table~\ref{tab:appendix-basis}).
The effect is similar on AIME, whose answers are also long, and small on
TeleQnA, whose answers are short and whose two clusters are large.

Table~\ref{tab:appendix-basis} therefore prices the TeleMath systems both ways.
Whichever way both are priced, CRE-Router at both TPOT budgets is comparable in
accuracy to the no-clustering ablation and costs less per token.
At $B = 25$\,ms, where its accuracy is comparable to Qwen3-30B-Thinking's,
CRE-Router costs 88 to 93\% of that model's TPOT when both are priced the same
way, and 74\% under the protocol of the main tables.

Both ways of pricing serve a fixed set of requests sent at once, as every
measurement in this paper does.
The gap between them shows how much the size of the workload can move a
comparison.
On either basis, CRE-Router remains cheaper per token than the no-clustering
ablation and than Qwen3-30B-Thinking at comparable accuracy; the protocol changes
the size of the saving, not its direction.

\section{AIME and TeleQnA in Full}
\label{sec:appendix-full}

Sections~\ref{sec:res-single} and~\ref{sec:res-baselines} summarise AIME and
TeleQnA. Table~\ref{tab:appendix-full} gives every system on both benchmarks and both
card counts: CRE-Router with and without escalation, each pool model alone, and
the two reference models.

\begin{table}[t]
\centering
\caption{\textbf{AIME and TeleQnA in full.} Every system on the AIME and TeleQnA test sets, on one and on two A100s: CRE-Router at Stage~1 and Stage~1+2 under the TPOT budget fitted for that card count, each pool model serving the whole test set alone, and the two reference models. Accuracy in \%, TPOT in ms, E2EL in s, means over runs, each request priced on the workload it was served in. Qwen3-Next-80B does not fit one A100.}
\label{tab:appendix-full}
\small
\begin{adjustbox}{max width=\textwidth}
\begin{tabular}{lrrrrrr}
\toprule
 & \multicolumn{3}{c}{\textbf{One A100}} & \multicolumn{3}{c}{\textbf{Two A100s}} \\
\cmidrule(lr){2-4}\cmidrule(lr){5-7}
\textbf{System} & \textbf{Acc.} & \textbf{TPOT} & \textbf{E2EL} & \textbf{Acc.} & \textbf{TPOT} & \textbf{E2EL} \\
\midrule
\multicolumn{7}{l}{\textit{AIME}, budget $B = 30$ / $20$\,ms} \\
CRE-Router, Stage~1+2 & 85.1 & 20.1 & 362.7 & 87.0 & 13.5 & 246.1 \\
CRE-Router, Stage~1 & 83.8 & 19.5 & 348.8 & 84.7 & 12.7 & 227.6 \\
VibeThinker-1.5B & 73.8 & 10.8 & 187.9 & 72.9 & 6.7 & 114.2 \\
Qwen3-30B-think & 84.9 & 37.0 & 666.9 & 84.4 & 22.9 & 406.6 \\
Qwen3-8B, reference & 70.4 & 42.0 & 711.0 & 69.8 & 23.1 & 404.8 \\
Qwen3-Next-80B-A3B-Thinking, reference & \multicolumn{3}{c}{--} & 88.0 & 19.0 & 357.5 \\
\midrule
\multicolumn{7}{l}{\textit{TeleQnA}, budget $B = 20$ / $15$\,ms} \\
CRE-Router, Stage~1+2 & 75.1 & 24.1 & 1.2 & 74.7 & 16.9 & 0.8 \\
CRE-Router, Stage~1 & 71.0 & 18.2 & 0.9 & 71.2 & 13.0 & 0.6 \\
Qwen3-4B-Instruct & 66.8 & 10.6 & 0.5 & 67.2 & 8.8 & 0.4 \\
Gemma4-E2B & 63.6 & 8.4 & 0.3 & 63.9 & 10.9 & 0.5 \\
Gemma4-E4B & 67.9 & 13.7 & 0.6 & 67.4 & 12.3 & 0.5 \\
Gemma4-26B & 76.6 & 30.0 & 1.5 & 76.6 & 19.6 & 1.0 \\
Qwen3-8B, reference & 73.4 & 15.8 & 14.1 & 73.3 & 9.7 & 8.6 \\
Qwen3-Next-80B-A3B-Instruct, reference & \multicolumn{3}{c}{--} & 72.7 & 36.4 & 2.0 \\
Qwen3-Next-80B-A3B-Thinking, reference & \multicolumn{3}{c}{--} & 76.7 & 17.6 & 24.6 \\
\bottomrule
\end{tabular}
\end{adjustbox}
\end{table}

\section{Cost of Stage 2}
\label{sec:appendix-stage2-cost}

Table~\ref{tab:stage2-cost} gives, for every reported routing, the share of
queries Stage~2 escalates and the TPOT and E2EL it adds to Stage~1 on the test
set, beside an estimate from training data alone.
An escalated query adds the strong model's E2EL, so the added E2EL grows with the
share escalated and with the strong model's cost.
The added TPOT also depends on output lengths (Equation~\ref{eq:tpot-esc}).
On TeleQnA, where the strong model's answers are short, it adds 4 to 6\,ms,
though only 0.2 to 0.3\,s per request; escalating to a reasoning model with long
answers adds little, and on TeleMath at $B = 120$\,s it lowers the TPOT slightly.
The training estimate assumes an estimator that escalates exactly the wrong
answers, at the efficient model's training error rate, since our estimators stop
early on the test answers and leave no held-out answers to measure their own rate
on.
On TeleMath at the three tighter budgets on one A100, the estimators escalate
about half of the wrong answers, with a precision above 80\%, so the estimate
overstates the added TPOT there.

An operator who needs a lower total cost has two options, lowering the budget,
so that Stage~1 selects a cheaper routing, or lowering the threshold $\tau$, so
that Stage~2 escalates fewer answers.
On TeleMath on one A100, at $B = 25$\,ms, escalating nothing saves 1.0\,ms of
TPOT and loses 2.7 points of accuracy.
Moving to $B = 20$\,ms instead saves 2.3\,ms and loses 3.2 points, although the
cheaper routing escalates twice as many queries.
Each millisecond saved thus costs 2.7 points through the threshold but 1.4
through the budget.
We therefore keep $\tau = 0.5$, the classifier's own decision, on every routing,
and set the cost with the budget.
A lower threshold pays only where Stage~2 adds little accuracy; at
$B = 120$\,s, $\tau = 0.2$ cuts E2EL by 6.7\,s and switching Stage~2 off by
21.0\,s, both at comparable accuracy.
Each lower threshold escalates a subset of the $\tau = 0.5$ escalations and is
priced on those batches as served.

\begin{table}[t]
\centering
\caption{\textbf{What Stage~2 adds to the budget.} For each reported routing, the share of queries escalated and the TPOT and E2EL that Stage~2 adds to Stage~1 on the test set, with 95\% intervals, beside an estimate from training data alone: the routing the budget selects from the training statistics, each gated cluster escalating its efficient model's wrong training answers, priced with the strong model's training cost. The budget $B$ bounds Stage~1; on every routing, the increment can be estimated from training data alone with an error of at most 7\% of the total E2EL.}
\label{tab:stage2-cost}
\small
\begin{adjustbox}{max width=\textwidth}
\begin{tabular}{lllrrrrr}
\toprule
 & & & \multicolumn{3}{c}{\textbf{Test set}} & \multicolumn{2}{c}{\textbf{Training estimate}} \\
\cmidrule(lr){4-6} \cmidrule(lr){7-8}
\textbf{Benchmark} & \textbf{GPUs} & \textbf{Budget} & \textbf{Escalated (\%)} & \textbf{$+$TPOT (ms)} & \textbf{$+$E2EL (s)} & \textbf{$+$TPOT (ms)} & \textbf{$+$E2EL (s)} \\
\midrule
TeleMath & one & TPOT 20\,ms & 16.1 & $+1.8$ {\scriptsize [$+1.2$, $+2.4$]} & $+21.3$ {\scriptsize [$+14.2$, $+29.3$]} & $+3.2$ & $+23.3$ \\
 & one & TPOT 25\,ms & 8.2 & $+1.0$ {\scriptsize [$+0.6$, $+1.5$]} & $+11.8$ {\scriptsize [$+6.3$, $+18.0$]} & $+2.1$ & $+13.6$ \\
 & one & E2EL 25\,s & 8.2 & $+2.3$ {\scriptsize [$+1.3$, $+3.6$]} & $+2.4$ {\scriptsize [$+1.1$, $+4.0$]} & $+3.1$ & $+3.3$ \\
 & one & E2EL 120\,s & 13.3 & $-0.3$ {\scriptsize [$-0.5$, $-0.1$]} & $+21.0$ {\scriptsize [$+13.4$, $+29.5$]} & $+0.3$ & $+22.5$ \\
 & two & TPOT 20\,ms & 21.8 & $+2.3$ {\scriptsize [$+1.7$, $+3.0$]} & $+59.1$ {\scriptsize [$+43.5$, $+75.4$]} & $+1.9$ & $+45.2$ \\
\midrule
TeleQnA & one & TPOT 20\,ms & 21.4 & $+5.9$ {\scriptsize [$+5.2$, $+6.6$]} & $+0.32$ {\scriptsize [$+0.28$, $+0.37$]} & $+5.0$ & $+0.25$ \\
 & two & TPOT 15\,ms & 21.8 & $+3.9$ {\scriptsize [$+3.4$, $+4.3$]} & $+0.22$ {\scriptsize [$+0.19$, $+0.25$]} & $+3.2$ & $+0.17$ \\
\midrule
AIME & one & TPOT 30\,ms & 4.2 & $+0.6$ {\scriptsize [$+0.2$, $+1.0$]} & $+13.9$ {\scriptsize [$+2.9$, $+28.7$]} & $+1.5$ & $+18.5$ \\
 & two & TPOT 20\,ms & 6.7 & $+0.8$ {\scriptsize [$+0.3$, $+1.5$]} & $+18.5$ {\scriptsize [$+5.3$, $+36.3$]} & $+1.0$ & $+11.8$ \\
\bottomrule
\end{tabular}
\end{adjustbox}
\end{table}

\section{Pool Size}
\label{sec:appendix-pool}

Restricting the pool to two models, as HybridLLM must
(Table~\ref{tab:telemath-main}), leaves the accuracy of CRE-Router comparable, with or
without escalation, but the full pool is significantly cheaper per token at both budgets, at both stages.
At $B = 25$\,ms the two-model routing can only send every question to
Qwen3-30B, which exceeds the budget on the test set, while the full pool stays within it.

\section{Hardware in Detail}
\label{sec:appendix-hardware}

Section~\ref{sec:res-hardware} reports what a second card changes.
Table~\ref{tab:appendix-hw-speed} gives the cost per token of each TeleMath pool
model on one and two A100s, Table~\ref{tab:appendix-hw-routings} the routing
each card count is reported with, and Table~\ref{tab:appendix-hw-samebudget} the
routing each selects at the same budget.

\begin{table}[H]
\centering
\caption{\textbf{A second card speeds models up unevenly.} Each TeleMath pool model's TPOT on the training set, as the routings are fitted on it, on one A100 and on two with tensor parallelism, and the speed-up, the ratio of the two. The speed-up ranges from 1.0 to 1.7, so a second card reorders the models' costs relative to one another.}
\label{tab:appendix-hw-speed}
\small
\begin{tabular}{lrrr}
\toprule
\textbf{Model} & \textbf{1$\times$ A100 (ms)} & \textbf{2$\times$ A100 (ms)} & \textbf{Speed-up} \\
\midrule
Qwen3-4B-think & 24.5 & 14.1 & 1.74 \\
Gemma4-26B-think & 33.3 & 19.3 & 1.72 \\
Gemma4-26B & 24.7 & 15.9 & 1.56 \\
Qwen3-30B-think & 24.6 & 16.2 & 1.52 \\
Qwen3-4B-Instruct & 12.4 & 8.3 & 1.51 \\
Gemma4-E4B-think & 12.4 & 9.5 & 1.30 \\
Gemma4-E4B & 11.8 & 9.0 & 1.30 \\
Gemma4-E2B & 7.6 & 7.1 & 1.07 \\
Gemma4-E2B-think & 8.1 & 7.9 & 1.02 \\
\bottomrule
\end{tabular}
\end{table}

\begin{table}[H]
\centering
\caption{\textbf{Routings reported on one and two A100s.} The routing CRE-Router selects on each benchmark on one and on two A100s, each card count fitted on its own measured costs under its own TPOT budget $B$, the routings of Table~\ref{tab:hardware}. A dagger marks a cluster Stage~2 gates, escalating to the model in the escalation column. The last column counts the A100s a routing loads, the models it uses times the cards per model. AIME has three clusters, TeleQnA two and TeleMath four.}
\label{tab:appendix-hw-routings}
\small
\begin{adjustbox}{max width=\textwidth}
\begin{tabular}{lllllllll}
\toprule
\textbf{Benchmark} & \textbf{GPUs} & \textbf{$B$ (ms)} & \textbf{C0} & \textbf{C1} & \textbf{C2} & \textbf{C3} & \textbf{Escalates to} & \textbf{A100s} \\
\midrule
TeleMath & one & 25 & Qwen3-30B-think & Gemma4-E2B$^\dagger$ & Qwen3-30B-think & Gemma4-26B-think & Qwen3-30B-think & 3 \\
 & two & 20 & Qwen3-30B-think$^\dagger$ & Qwen3-30B-think$^\dagger$ & Gemma4-26B-think & Gemma4-26B-think & Gemma4-26B-think & 4 \\
\midrule
TeleQnA & one & 20 & Qwen3-4B-Instruct$^\dagger$ & Gemma4-26B & -- & -- & Gemma4-26B & 2 \\
 & two & 15 & Qwen3-4B-Instruct$^\dagger$ & Gemma4-26B & -- & -- & Gemma4-26B & 4 \\
\midrule
AIME & one & 30 & Qwen3-30B-think & VibeThinker-1.5B$^\dagger$ & Qwen3-30B-think & -- & Qwen3-30B-think & 2 \\
 & two & 20 & Qwen3-30B-think & VibeThinker-1.5B$^\dagger$ & Qwen3-30B-think & -- & Qwen3-30B-think & 4 \\
\bottomrule
\end{tabular}
\end{adjustbox}
\end{table}

\begin{table}[H]
\centering
\caption{\textbf{At the same budget, the hardware changes the routing.} The routing CRE-Router selects on one and on two A100s at every TPOT budget $B$ reported for either card count, each card count fitted on its own measured costs. A dagger marks the routing a card count is reported with (Table~\ref{tab:hardware}). At every budget the two card counts select different routings.}
\label{tab:appendix-hw-samebudget}
\small
\begin{adjustbox}{max width=\textwidth}
\begin{tabular}{lllllll}
\toprule
\textbf{Benchmark} & \textbf{$B$ (ms)} & \textbf{GPUs} & \textbf{C0} & \textbf{C1} & \textbf{C2} & \textbf{C3} \\
\midrule
TeleMath & 25 & one$^\dagger$ & Qwen3-30B-think & Gemma4-E2B & Qwen3-30B-think & Gemma4-26B-think \\
 &  & two & Qwen3-30B-think & Qwen3-30B-think & Gemma4-26B-think & Gemma4-26B-think \\
 & 20 & one & Qwen3-30B-think & Gemma4-E2B & Qwen3-30B-think & Gemma4-E4B-think \\
 &  & two$^\dagger$ & Qwen3-30B-think & Qwen3-30B-think & Gemma4-26B-think & Gemma4-26B-think \\
\midrule
TeleQnA & 20 & one$^\dagger$ & Qwen3-4B-Instruct & Gemma4-26B & -- & -- \\
 &  & two & Gemma4-26B & Gemma4-26B & -- & -- \\
 & 15 & one & Qwen3-4B-Instruct & Qwen3-4B-Instruct & -- & -- \\
 &  & two$^\dagger$ & Qwen3-4B-Instruct & Gemma4-26B & -- & -- \\
\midrule
AIME & 30 & one$^\dagger$ & Qwen3-30B-think & VibeThinker-1.5B & Qwen3-30B-think & -- \\
 &  & two & Qwen3-30B-think & Qwen3-30B-think & Qwen3-30B-think & -- \\
 & 20 & one & VibeThinker-1.5B & VibeThinker-1.5B & VibeThinker-1.5B & -- \\
 &  & two$^\dagger$ & Qwen3-30B-think & VibeThinker-1.5B & Qwen3-30B-think & -- \\
\bottomrule
\end{tabular}
\end{adjustbox}
\end{table}

\section{TeleMath on Two A100s}
\label{sec:appendix-two-card}

\begin{figure}[!htb]
\centering
\includegraphics[width=\linewidth]{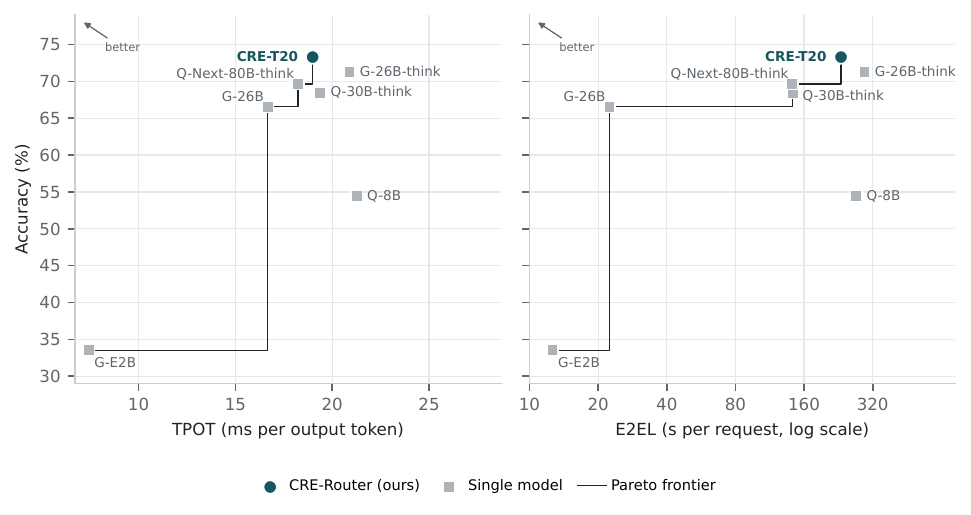}
\caption{\textbf{Pareto frontier of accuracy and cost on TeleMath, two A100s.} The two-card counterpart of Figure~\ref{fig:telemath-pareto}: CRE-Router at Stage~1+2 under its two-card TPOT budget, and every single model with a whole-test-set run on two A100s, Qwen3-Next-80B-Thinking among them, which does not fit on one. No prior method or E2EL routing was run on two cards. Model labels abbreviate Gemma4 as G and Qwen3 as Q. Higher and further left is better; the line is the Pareto frontier.}
\label{fig:telemath-pareto-2x}
\end{figure}

Figure~\ref{fig:telemath-pareto-2x} places the two-card TeleMath system of
Table~\ref{tab:hardware} among the single models served on two A100s.

\section{Concurrency in Detail}
\label{sec:appendix-concurrency}

Table~\ref{tab:concurrency} gives the values behind Figure~\ref{fig:concurrency}.
Throughput rises with load for both models.
At 32 requests the whole-run figure vLLM reports is close to the full-load
throughput for the dense model but well below it for the mixture of experts,
because the whole run includes its end, when only a few long
answers are still being generated and most slots are idle.
We report both.

\begin{table}[!htb]
\centering
\caption{\textbf{Effect of concurrency.} Cost and throughput against load{,} for two models serving the TeleMath test set without reasoning: Gemma4-E4B{,} a dense 4B-effective model{,} and Gemma4-26B{,} a mixture of experts with 4B active parameters. Two runs per point. (a) TPOT at each number of concurrent requests. (b) Output throughput at 32 concurrent requests{,} over the whole run as vLLM reports it{,} and at full load{,} while every slot is busy. On one A100 the two models cost about the same per token at one request{,} and at 32 the mixture of experts costs about twice as much. Every other result in the paper is measured at 32.}
\label{tab:concurrency}
\small
\begin{tabular}{rrrrr}
\toprule
 & \multicolumn{2}{c}{\textbf{1$\times$ A100}} & \multicolumn{2}{c}{\textbf{2$\times$ A100}} \\
\cmidrule(lr){2-3}\cmidrule(lr){4-5}
\textbf{Concurrent requests} & \textbf{Gemma4-E4B} & \textbf{Gemma4-26B} & \textbf{Gemma4-E4B} & \textbf{Gemma4-26B} \\
\midrule
\multicolumn{5}{l}{\textit{(a) TPOT (ms)}} \\
1 & 9.3 & 9.1 & 6.8 & 7.4 \\
2 & 9.9 & 11.5 & 7.8 & 9.0 \\
4 & 10.5 & 14.1 & 8.2 & 10.6 \\
8 & 10.6 & 16.9 & 8.1 & 12.0 \\
16 & 11.3 & 20.2 & 8.7 & 14.4 \\
32 & 12.4 & 26.5 & 9.5 & 16.7 \\
\midrule
\multicolumn{5}{l}{\textit{(b) Output throughput at 32 (tokens/s)}} \\
Whole run & 2{,}297 & 699 & 2{,}892 & 991 \\
Full load & 2{,}542 & 1{,}177 & 3{,}325 & 1{,}875 \\
\bottomrule
\end{tabular}
\end{table}

\section{Overhead of Routing and Estimation}
\label{sec:appendix-overhead}

\begin{table}[H]
\centering
\caption{\textbf{Routing and estimation overhead.} Time added by the router and the quality estimator on one A100, median milliseconds per request, not per token: the router runs once per query and the estimator once per gated answer, alone (batch 1) or in batches of 32. The last two columns put the batch-1 time in the units of the two cost terms, as a share of the mean E2EL of CRE-Router on that benchmark and spread over its mean output tokens per request, estimated as E2EL divided by TPOT. Every component adds at most 1.4\% to E2EL, the estimator on TeleQnA, whose answers are short.}
\label{tab:overhead}
\small
\begin{adjustbox}{max width=\textwidth}
\begin{tabular}{llrrrr}
\toprule
\textbf{Component} & \textbf{Benchmark} & \textbf{Batch 1 (ms)} & \textbf{Batch 32 (ms)} & \textbf{E2EL share (\%)} & \textbf{Per token (ms)} \\
\midrule
Stage~1 router & TeleMath & 4.5 & 0.9 & 0.0023 & 0.00049 \\
Stage~2 estimator & TeleMath & 19.8 & 6.2 & 0.010 & 0.0022 \\
 & TeleQnA & 16.3 & 1.0 & 1.4 & 0.33 \\
 & AIME & 20.8 & 3.4 & 0.0057 & 0.0012 \\
\bottomrule
\end{tabular}
\end{adjustbox}
\end{table}

Table~\ref{tab:overhead} gives the times behind Section~\ref{sec:res-overhead},
measured on one A100, alone and in batches of 32.
These are times per request.
Spread over the output tokens of a request, the estimator adds 0.33\,ms per token
on TeleQnA and less than 0.003\,ms on the reasoning benchmarks.
As the token count is estimated as E2EL divided by TPOT, each component's share of
TPOT equals its share of E2EL.

\end{document}